\documentclass[longauth,singlecolumn]{aa}
\usepackage[varg]{txfonts}
\usepackage{graphics}
\usepackage{epstopdf}
\usepackage{hyperref}
\usepackage[x11names]{xcolor}
\usepackage{enumerate}
\usepackage{subfig}
\renewcommand{\textbf}{\textnormal}
\usepackage{array}

\usepackage{natbib}
\begin{document}

\captionsetup[figure]{labelformat=simple, labelsep=period}
\captionsetup[table]{labelformat=simple, labelsep=period}

\newcommand{\Msun}{\rm{M}_\odot}
\newcommand{\msun}{\rm{M}_\odot}
\newcommand{\mste}{\rm{M}_{\star}}
\newcommand{\mstar}{\rm{M}^{\star}}
\newcommand{\rtwo}{\rm{r}_{200}}
\newcommand{\ABi}[1]{\textcolor{blue}{\textbf{Andrea: #1}}}

\title{CLASH-VLT: The stellar mass function and stellar mass density
  profile of the z=0.44 cluster of galaxies MACS~J1206.2-0847
  \thanks{Based in large part on data collected at the ESO VLT
    (prog.ID 186.A-0798), at the NASA HST, and at the NASJ Subaru
    telescope}}

\author{M. Annunziatella\inst{\ref{MAn},\ref{ABi}}
\and A. Biviano\inst{\ref{ABi}}
\and A. Mercurio\inst{\ref{AMe}} 
\and M. Nonino\inst{\ref{ABi}} 
\and P. Rosati\inst{\ref{PRo}}
\and I. Balestra\inst{\ref{ABi}}
\and V. Presotto\inst{\ref{MAn},\ref{ABi}}
\and M. Girardi\inst{\ref{MAn},\ref{ABi}} 
\and R. Gobat\inst{\ref{RGo}} 
\and C. Grillo\inst{\ref{CGr}}
\and D. Kelson\inst{\ref{DKe}}
\and E. Medezinski\inst{\ref{EMe}}
\and M. Postman\inst{\ref{MPo}}
\and M. Scodeggio\inst{\ref{MSc}}
\and M. Brescia\inst{\ref{AMe}}
\and R. Demarco\inst{\ref{RDe}}
\and A. Fritz\inst{\ref{MSc}}
\and A. Koekemoer\inst{\ref{MPo}}
\and D. Lemze\inst{\ref{EMe}}
\and M. Lombardi\inst{\ref{MLo}}
\and B. Sartoris\inst{\ref{MAn},\ref{ABi},\ref{BSa}}
\and K. Umetsu\inst{\ref{KUm}}
\and E. Vanzella\inst{\ref{EVa}}
\and L. Bradley\inst{\ref{MPo}}
\and D. Coe\inst{\ref{MPo}}
\and M. Donahue\inst{\ref{MDo}}
\and L. Infante\inst{\ref{LIn}}
\and U. Kuchner\inst{\ref{BZi}}
\and C. Maier\inst{\ref{BZi}}
\and E. Reg\H{o}s\inst{\ref{ERe}}
\and M. Verdugo\inst{\ref{BZi}}
\and B. Ziegler\inst{\ref{BZi}}
}

\offprints{M. Annunziatella, annunziatella@oats.inaf.it}

\institute{Dipartimento di Fisica, Univ. degli Studi di Trieste, via Tiepolo 11, I-34143 Trieste, Italy\label{MAn} \and
INAF/Osservatorio Astronomico di Trieste, via G. B. Tiepolo 11, 
I-34131, Trieste, Italy\label{ABi} \and
INAF/Osservatorio Astronomico di Capodimonte, Via Moiariello 16 I-80131 Napoli, Italy\label{AMe} \and
Dipartimento di Fisica e Scienze della Terra, Univ. degli Studi di Ferrara, via Saragat 1, I-44122, Ferrara, Italy\label{PRo} \and
Laboratoire AIM-Paris-Saclay, CEA/DSM-CNRS, Universit\'e Paris Diderot, Irfu/Service d'Astrophysique, CEA Saclay, Orme des Merisiers, F-91191 Gif sur Yvette, France\label{RGo} \and
Dark Cosmology Centre, Niels Bohr Institute, University of Copenhagen,
Juliane Maries Vej 30, 2100 Copenhagen, Denmark\label{CGr} \and 
Observatories of the Carnegie Institution of Washington, Pasadena, CA 91 101, USA\label{DKe} \and
Department of Physics and Astronomy, The Johns Hopkins University, 3400 North Charles Street, Baltimore, MD 21218, USA\label{EMe} \and
Space Telescope Science Institute, 3700 San Martin Drive, Baltimore, MD 21218, USA\label{MPo} \and
INAF/IASF-Milano, via Bassini 15, 20133 Milano, Italy\label{MSc} \and
Department of Astronomy, Universidad de Concepci\'on, Casilla 160-C, Concepci\'on, Chile \label{RDe} \and
Dipartimento di Fisica, Universit\`a degli Studi di Milano, via Celoria 16, I-20133 Milan, Italy\label{MLo} \and
INFN, Sezione di Trieste, via Valerio 2, I-34127 Trieste, Italy\label{BSa} \and
Institute of Astronomy and Astrophysics, Academia Sinica, P. O. Box 23-141, Taipei 10617, Taiwan\label{KUm}
INAF/Osservatorio Astronomico di Bologna, via Ranzani 1, I-40127, Bologna, Italy\label{EVa} \and
Physics and Astronomy Dept., Michigan State University, 567 Wilson Rd., East Lansing, MI 48824, USA\label{MDo} \and
Pontificia Universidad Cat\'olica de Chile, Departamento de Astronom\'{\i}a y Astrof\'{\i}sica, Av. Vicu\~na Mackenna 4860, Santiago, Chile\label{LIn} \and
University of Vienna, Department of Astrophysics, T\"urkenschanzstr. 17, 1180 Wien, Austria\label{BZi} \and
European Laboratory for Particle Physics (CERN), CH-1211, Geneva 23, Switzerland\label{ERe}
}

\date{}

\abstract{The study of the galaxy stellar mass function (SMF) in
  relation to the galaxy environment and the stellar mass density
  profile, $\mathrm{\rho_{\star}(r)}$,is a powerful tool to constrain
  models of galaxy evolution.}{ We determine the SMF of the z=0.44
  cluster of galaxies MACS~J1206.2-0847 separately for passive and
  star-forming (SF) galaxies, in different regions of the cluster,
  from the center out to approximately 2 virial radii. We also
  determine $\mathrm{\rho_{\star}(r)}$ to compare it to the number
  density and total mass density profiles.  }{We use the dataset from
  the CLASH-VLT survey. Stellar masses are obtained by spectral energy
  distribution fitting with the \texttt{MAGPHYS} technique on 5-band
  photometric data obtained at the Subaru telescope. We identify 1363
  cluster members down to a stellar mass of $10^{9.5} \msun$, selected
  on the basis of their spectroscopic ($\sim 1/3$ of the total) and
  photometric redshifts. We correct our sample for incompleteness and
  contamination by non members. Cluster member environments are
  defined using either the clustercentric radius or the local galaxy
  number density. }{The whole cluster SMF is well fitted by a 
  double Schechter function, which is the sum of the two Schechter
    functions that provide good fits to the SMFs of, separately, the
  passive and SF cluster populations. The SMF of SF galaxies is
    significantly steeper than the SMF of passive galaxies at the
    faint end.  The SMF of the SF cluster galaxies does not depend on
  the environment. The SMF of the passive cluster galaxies has a
  significantly smaller slope (in absolute value) in the innermost
  ($\leq 0.50$ Mpc, i.e., $\sim 0.25$ virial radii), and in the highest
  density cluster region than in more external, lower density
  regions. The number ratio of giant/subgiant galaxies is maximum in
  this innermost region and minimum in the adjacent region, but then
  gently increases again toward the cluster outskirts. This is also
  reflected in a decreasing radial trend of the average stellar mass
  per cluster galaxy. On the other hand, the stellar mass fraction,
  i.e., the ratio of stellar to total cluster mass, does not show any
  significant radial trend. }{Our results appear consistent with a
  scenario in which SF galaxies evolve into passive galaxies due to
  density-dependent environmental processes, and eventually get
  destroyed very near the cluster center to become part of a diffuse
  intracluster medium. Dynamical friction, on the other hand, does
  not seem to play an important role. Future investigations of other
  clusters of the CLASH-VLT sample will allow us to confirm our
  interpretation.}


\titlerunning{CLASH cluster stellar mass function}
\authorrunning{M. Annunziatella et al.}

\maketitle

\section{Introduction}
\label{s:intro}
Many galaxy properties, such as
colors, luminosities, morphologies, star formation rates, and stellar
masses, follow a bimodal distribution (e.g., \citealt{baldry2004,kauffmann2003}).  Galaxies can
therefore be classified in two broad classes, red, bulge-dominated,
high-mass, passively-evolving galaxies, and blue, disk-dominated,
low-mass, star-forming galaxies.  The relative number fraction of
these two populations changes with redshift (z) and with the local
galaxy number density, blue galaxies dominating at higher z and in
lower density environments (see \citealt{silk&manon2012} for a recent
review on galaxy formation and evolution).  This suggests that the
redshift evolution of these two populations is somehow shaped by
  physical processes related to the environment in which they reside,
  such as major and minor mergers, tidal interactions among galaxies
or between a galaxy and a cluster gravitational field, and
ram-pressure stripping (see, e.g., \citealt{biviano2008} and
references therein).  All these processes use or remove gas from
  galaxies, leading to a drop in star-formation due to lack
of fuel and to an aging of the stellar population, and consequent
reddening of the galaxy light with time. Some of these processes also
lead to morphological transformations. These quenching mechanisms have
been implemented in both N-body simulations and semi-analytical models
with the aim of reproducing the phenomenology of galaxy evolution, and
in particular the changing fraction of red and blue galaxies with
time.  However, there are still many discrepancies between
observations and theoretical predictions,  such as, e.g., the
  evolution of galaxy colors, luminosities, and stellar masses
(e.g., \citealt{cucciati2012}; \citealt{delucia2012};
\citealt{silk&manon2012} and references therein).

The distributions of galaxy luminosities and stellar masses ($\mste$
hereafter), namely, the galaxy luminosity and stellar mass functions
(SMF hereafter), are key observables for testing galaxy evolutionary
models (e.g., \citealt{maccio2010,menci2012}).  The SMF allows for a
more direct test of theoretical models than the luminosity function,
since luminosities are more difficult to predict than $\mste$ because of
effects such as the age and metallicity of the stellar population, the
dust content of the interstellar medium, etc. On the other hand,
  unlike luminosities, $\mste$ are not direct observables, and
  can be determined only via multicolor and/or near infrared
photometry.  This explains why most studies of the galaxy SMF have
been conducted only quite recently.

Most determinations of the galaxy SMF (or of the near-infrared
luminosity function, which is considered a proxy for the SMF) have
been based on samples of field galaxies. The field galaxy SMF
appears to have a flat slope down to $10^9 \msun$, up to $\mathrm{z} \sim 1$
\citep{fontana2006} and beyond
(\citealp{stefanon&marchesini2013}; \citealt{sobral2014}),
although some authors provide evidence
that the SMF steepens with z (\citealt{mortlock2011};
	\citealt{bielby2012}; \citealt{huang2013}).
\cite{ilbert2010} find that
%that the SMF of SF galaxies varies little with z, but that of 
%passive galaxies steepens considerably with time.
this steepening occurs at masses lower than a certain limit,
which varies with z, and results from the combination
of two single \cite{schechter1976} functions
that characterize, separately, the red and blue SMF
(\citealt{bolzonella2010}; \citealt{ilbert2010};
\citealt{pozzetti2010}).

%a double Schechter function is required
%(\citealt{baldry2008}, \citealt{drory2009}). 

%but the red SMF itself appears to be
%inconsistent with a single Schechter, due to an upturn at
%small masses. The full galaxy SMF might however be even more
%complicated, since within the two broad color classes of galaxies,
%ellipticals and S0s, on one side, and early- and late-spirals, on the
%other side, have different SMFs (\citealt{bernardi2013}).

To highlight possible environmental effects on the galaxy SMF one
should compare the field galaxy SMF to that of cluster galaxies.
\cite{balogh2001} have found the SMF of non emission line
galaxies to be steeper in clusters than in the field.  On the
other hand, \cite{vulcani2012, vulcani2013} have found the field
and cluster SMF not to be different, at least down to $\mste \sim
10^{10.2} \msun$, not even when considering different galaxy
  populations separately. Their analysis is based on optical
magnitudes and colors, while \cite{balogh2001} use J-band
magnitudes. Other studies of the near infrared luminosity
functions of cluster and field galaxies found them to be 
  statistically indistinguishable (\citealt{lin2004};
\citealt{strazzullo2006}; and
\citealt{depropris2009}). \cite{giodini2012} find no major difference
between the SMF of field and group SF galaxies, at any redshift, and
down to $\sim 10^{8.5} \msun$, except at the high-mass end, however,
they do find significant differences in the SMF of passive galaxies in
the field and low-mass groups, on one side, and in high-mass groups,
on the other. Within clusters, there is no difference in the global
SMFs evaluated within and outside the virial region
(\citealt{vulcani2013}), but \cite{calvi2013} find the SMFs of
different galaxy types change within different cluster
  environments.

Different results might be caused by the different $\mste$
completeness limits reached by the different studies.
\cite{merluzzi2010} suggest that at low z the environmental dependence
of the SMF becomes evident only for masses below $\sim 10^9 \,
\msun$. At $\mathrm{z} \sim 1$ an environmental dependence of the
  SMF is already seen at the $10^{10} \, \msun$ mass limit
\citep{vanderburg2013}. This mass limit may in fact depend on
redshift, as it corresponds to the mass below which the relative
  contribution of blue galaxies to the SMF becomes dominant
  \cite{Davidzon+13}. In fact, red galaxies show in fact a milder evolution
  with z than blue galaxies (at least for masses $\geq 10^{11.4}
  \msun$ \citealt{Davidzon+13}) and they are more abundant in denser
  environments, at least until $\mathrm{z} \simeq 1.5$.

The SMF massive end,  dominated by red
  galaxies, seems to be already in place at high z
(\citealt{kodama&bower2003}; \citealt{andreon2013}) in clusters, and the
characteristic magnitude of the near-infrared luminosity function of
cluster galaxies evolves as predicted by models of passive stellar
evolution (\citealt{lin2006}; \citealt{strazzullo2006};
\citealt{depropris2007}; \citealt{muzzin2007, muzzin2008};
\citealt{capozzi2012}; \citealt{mancone2012}). \cite{mancone2012} 
do not detect any evolution of the slope of the near-infrared
luminosity function of clusters up to $\mathrm{z} \sim 1.5$,  but
  their result appears to contrast with the claimed evolution of the
slope of the cluster SMF from $\mathrm{z} \sim 0$ to 0.5 by
\cite{vulcani2011}.

\begin{table}[ht]
\centering
\caption{Main properties of the cluster MACS~J1206.2-0847}
\label{t:props}
%\small
% \begin{tabular}{p{4cm} p{2cm} p{2cm}}
\begin{tabular}{l>{\bfseries}c>{\bfseries}c>{\bfseries}lc}
 \hline
Center $(\alpha,\delta)_{\mathrm{J2000}}$ & $12^{\mathrm{h}}06^{\mathrm{m}}12\fs15, -8\degr 48\arcmin 3\farcs4$ \\
Mean redshift & $0.43984 \pm 0.00015$\\
Velocity dispersion [km~s$^{-1}$] & $1087_{-55}^{+53}$ \\
Virial radius $\mathrm{r_{200}}$ [Mpc] & $1.96 \pm 0.11$ \\
Virial mass $\mathrm{M_{200}}$ [$10^{15} \mathrm{M}_{\odot}$] & $1.37 \pm 0.23$ \\
\hline
\end{tabular}
\tablefoot {All values from \citealt{biviano2013}.}
\end{table}

In this paper, we determine the SMF of galaxies in the
$\mathrm{z}=0.44$ cluster MACS~J1206.2-0847 \citep[M1206 hereafter),
  discovered by][]{ebeling2009,Ebeling+01}, and part of the CLASH
(``Cluster Lensing And Supernova survey with Hubble'') sample
(\citealt{postman2012}). We provide the main properties of this cluster
in Table~\ref{t:props}.
We consider passive and SF cluster
members separately, and examine the dependence of their SMFs on the local density
and clustercentric radius, in a very wide radial range, $0-6$ Mpc
from the cluster center. This cluster has a unique spectroscopic
dataset of $\sim 600$ cluster members with redshifts measured with
VLT/VIMOS (\citealt{biviano2013,Lemze2013}). This dataset allows us to base our
SMF determination on a sample with a large fraction ($\sim 1/3$) of
spectroscopically confirmed (and hence secure) cluster members down to
$\mste=10^{9.5} \msun$. High quality five band photometry obtained with
the Subaru telescope provides photometric redshifts for the rest of the
sample, the quality of which is improved thanks to the large spectroscopic
dataset available for calibration ($\sim 2000$ objects; see
Mercurio et al., in prep.).

The structure of this paper is the following. In Sect.~\ref{s:data}, we
describe the data sample, how we determine the cluster membership and
$\mste$ of galaxies in our data sample, and how we correct for
incompleteness and contamination. In Sect.~\ref{s:smf}, we describe how
we determine and model fit the cluster SMF, and examine the dependence
of the SMF from the galaxy type, the clustercentric radius, and the
local galaxy number density. In Sect.~\ref{s:smfrac}, we determine the
stellar mass density profile of our cluster and compare it to the
galaxy number density profile and the total mass density profile.  In
Sect.~\ref{s:disc}, we discuss our results. Finally, in
Sect.~\ref{s:conc} we summarize our results and draw our conclusions.

Throughout this paper, we use $\mathrm{H_0 \, = \, 70}$,
$\mathrm{\Omega_M \, = \, 0.3, }$ and $\mathrm{\Omega_{\Lambda} \, =
  \, 0.7 }$.

\section{The data sample}
\label{s:data}
We observed the cluster M1206 in 2012 as part of the ESO Large
  Programme ``Dark Matter Mass Distributions of Hubble Treasury
  Clusters and the Foundations of $\Lambda$CDM Structure Formation
  Models'' (P.I. Piero Rosati). We used VIMOS \citep{LeFevre+03} at
  the ESO VLT, with 12 masks (eight in low resolution and four in
  medium resolution), each with an exposure time of either 3 or $4
  \times 15$ minutes (10.7 hours in total). Data were reduced with
  VIPGI \citep{Scodeggio+05}. We obtained no redshift measurement for
  306 spectra. For the other 3240 spectra, we quantified the
  reliability of the redshift determinations based on repeated
  measurements.  For 2006 of the spectra, the estimated probability that they
  are correct is $> 92$ \%, and for another 720 it is 75 \%. We do not
  consider the remaining 514 lower quality redshifts in our analysis.
  We finally added to our sample another 68 reliable redshifts from
  the literature \citep{Lamareille+06,Jones+04,Ebeling+09} and from
  IMACS-GISMO observations at the Magellan telescope (Dan Kelson,
  private communication). Our final dataset contains 2749 objects
  with reliable redshift estimates, of which 2513 have $z>0$. From
  repeated measurements, we estimate the average error on the radial
  velocities to be 75 (resp. 153) km~s$^{-1}$ for the spectra
  observed with the medium resolution (resp. low resolution) grism.
  Full details on the spectroscopic sample observations and
  data reduction will be given in Rosati et al. (in prep.).

We retrieved raw Suprime-Cam data from
SMOKA\footnote{http://smoka.nao.ac.jp} \citep{Baba+02} in the
BV$\mathrm{R_C I_Cz'}$ bands and processed them as described in
\citet{umetsu2012}.  We obtained aperture corrected magnitudes in each
band and we used these magnitudes to derive photometric
redshifts, $\mathrm{z_{phot}}$, using a neural network method
\citep{brescia2013}. More details on the measurement of
$\mathrm{z_{phot}}$ can be found in \cite{biviano2013}, while a full
description of the method will be given in Mercurio et al. (in prep.).
This method is considered reliable down to $\mathrm{R_c} \, =\, 25.0$.
We use the AB magnitude system throughout this paper.

Since our spectroscopic sample is not complete, we need to rely in
part on the sample of galaxies with $\mathrm{z_{phot}}$.  We
considered only objects in the magnitude range $\mathrm{18\, \leq \,
  R_C\, \leq \, 24}$ to maximize the number of objects with
spectroscopic redshifts. Cluster membership for the galaxies
with z has been established using the `Clean' algorithm of
\citet[][see also \citealt{biviano2013}]{mamon2013}. This
  algorithm starts from a first guess of the cluster mass derived from
  a robust estimate of the cluster line-of-sight velocity dispersion
  $\sigma_{\mathrm{los}}$ via a scaling relation. This mass guess is
  used to infer the concentration of the cluster mass profile, assumed
  to be NFW \citep{NFW97}, from a theoretical mass concentration
  relation \citep{MDvdB08}. Given the mass and concentration of the
  cluster, and adopting the velocity anisotropy profile model of
  \citet{MBM10}, a theoretical $\sigma_{\mathrm{los}}$-profile is
  predicted and used to reject galaxies with rest-frame velocities
  outside $\pm 2.7 \, \sigma_{\mathrm{los}}\mathrm{(R)}$ at any clustercentic
  distance R. The procedure is iterated until convergence.

\begin{figure}[ht]
 \centering
\includegraphics[width=1.\columnwidth]{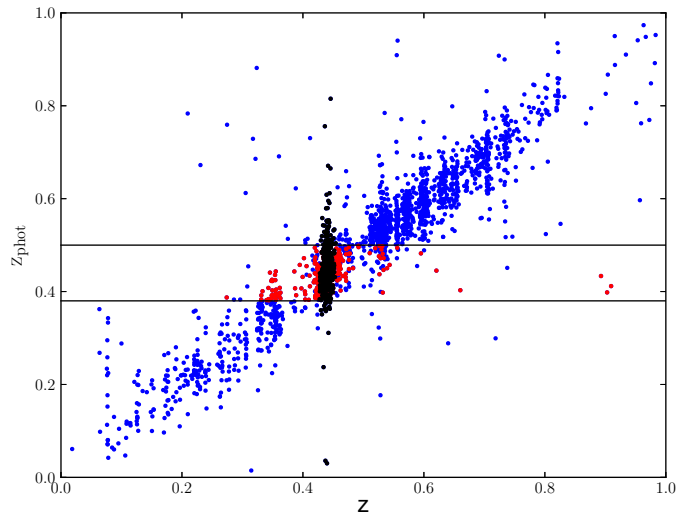}
\caption{Photometric vs. spectroscopic redshifts for galaxies in the
  cluster field and in the magnitude range $\mathrm{18\, \leq \, R_C\,
    \leq \, 24}$. Black dots represent spectroscopically confirmed
  members. The two horizontal lines indicate the $\mathrm{z_{phot}}$
  range chosen for membership selection of the galaxies without
  z. Within this range only galaxies with chosen colors are selected
  as members. In this diagram these galaxies are indicated as red
  dots. Blue crosses are galaxies outside the spectroscopical and
  photometrical membership selections. }
\label{f:zzp}
\end{figure}

Cluster membership for the galaxies without z, but with
$\mathrm{z_{phot}}$ has been obtained by investigation of the
$\mathrm{z_{phot}}$ vs. z diagram (see Fig.~\ref{f:zzp}), as described
in \citealt{biviano2013}.  In this diagram, we use the sample of
  galaxies with spectroscopic redshifts to investigate the best
  strategy for the selection of members among the sample without
  spectroscopic redshift.  In other words, we take advantage of our
  previous definition of cluster members with the `Clean' method to
  define cuts in $\mathrm{z_{phot}}$ and in colors that maximize the
  inclusion of cluster members and minimize that of interlopers. As it
  is evident from Fig.~\ref{f:zzp}, one cannot use too broad a range
  in $\mathrm{z_{phot}}$ for membership selection, or many foreground
  and background galaxies (the colored dots in Fig.~\ref{f:zzp}) would
  enter the sample of cluster members. On the other hand, trying to
  get rid of all foreground and background contamination would reject
  too many real members (the black dots in Fig.~\ref{f:zzp}).  As a
  compromise between these two extremes, of all galaxies without a
  spectroscopic redshift determination, we select those with
$0.38 < \mathrm{z_{phot}} < 0.50$ and within the $\mathrm{R_C-I_C}$
vs. $\mathrm{B-V}$ color cuts given in \cite{biviano2013}. Combining the samples of spectroscopically- and
  photometrically-selected members we obtain a sample of 2468 members
of which 590 are spectroscopically confirmed.

Unlike the spectroscopic selection of cluster members, the
  photometric selection is not secure. As seen in
  Fig.~\ref{f:zzp}, many galaxies selected as members based on their
  $\mathrm{z_{phot}}$ do not lie at the cluster spectroscopic
  redshift. We correct for this effect in Sect.~\ref{ss:compl}.

\subsection{Estimation of stellar mass}
\label{ss:mass}
Stellar masses of cluster member galaxies have been obtained, using the
spectral energy distribution (SED) fitting technique performed by
\texttt{MAGPHYS} (\citealt{dacunha2008}), by setting all member
galaxies to the mean cluster redshift. \texttt{MAGPHYS} uses a
Bayesian approach to choose the template that best reproduces
the observed galaxy SED. It is based on the stellar population
synthesis models of either \cite{bruzual&charlot2003} or Bruzual \&
Charlot (2007), with a \cite{chabrier2003} stellar initial mass
function and a metallicity value in the range 0.02--2
$\mathrm{Z_{\odot}}$.  The difference between the two libraries of
models is in the treatment of the thermally pulsating asymptotic giant
branch stellar phase, which affects the NIR emission of stellar
populations with an age of $\mathrm{\sim}$ 1 Gyr. There is still
considerable ongoing discussion on the way to model this phase of
stellar evolution (\citealt{maraston2006,kriek2010}).  We therefore
tried adopting both libraries and found no significant difference (on
average) in the stellar mass estimates. For simplicity, in the rest of
the paper we report results based only on the more traditional library
of models of \cite{bruzual&charlot2003}.

The spectral energy distribution is then obtained considering the 
  history of the star formation rate (SFR) parametrized as a continuum
model, $\mathrm{SFR \propto e^{-\gamma t}}$, with superimposed random
bursts. The timescale $\mathrm{\gamma}$ is distributed according to
the probability density function $\mathrm{p(\gamma) =1\, -\,
  \tanh(8\gamma \, - \, 6)}$, which is uniform between 0 and 0.6
$\mathrm{Gyr^{-1}}$ and drops exponentially to zero at 1
$\mathrm{Gyr^{-1}}$. In this model, the age of the galaxy is a free
parameter uniformly distributed over the interval from 0.1 to at most
13.5 Gyr. However, an upper limit for this value is provided by the
age of the universe at the considered redshift.

For each galaxy model MAGPHYS produces both the dust free and the attenuated spectrum. The attenuated spectra are obtained using the dust model of \citet{CF00}.  The main parameter of this model is  the total effective V-band absorption optical depth of the dust as seen by young stars inside birth clouds, $\mathrm{\widehat{\tau_V}}$. This parameter is distributed according to a probability density function which is approximately uniform over the interval from 0 to 4 and drops exponentially to zero at $\mathrm{\widehat{\tau_V}}\, \sim \, 6$.

As an output of the SED fitting procedure, \texttt{MAGPHYS} provides
both the parameters of the best-fit model and the marginalized
probability distribution of each parameter. We adopt the median value
of the probability distribution as our fiducial estimate of a given
parameter, with lower and upper limits provided by the 16\% and 84\%
percentiles of the same distribution. Using these limits we find that
the typical 1 $\sigma$ error on the $\mste$ estimates is $\sim 0.15$
dex. 

\begin{figure}[ht]
 \centering
\includegraphics[width=1.\columnwidth]{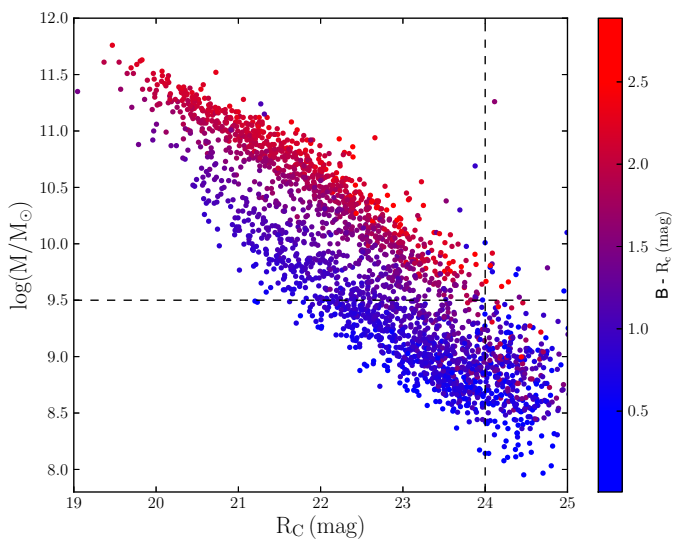}
\caption{Galaxy stellar mass as function of $\mathrm{R_C}$
    magnitude for cluster members.  The points are color coded
    according to their $\mathrm{B\,-\,R_C}$ color. The vertical dashed line
    represents the completeness magnitude of our sample, and the horizontal
    dashed line represents the corresponding completeness mass.}
\label{f:mr}
\end{figure}

We translate our completeness limit in magnitude,
  $\mathrm{R_C\, =\, 24}$, to a completeness limit in mass,
  $\mathrm{10^{9.5} \, \Msun}$, based on the relation between these two
  quantities shown in Fig.~\ref{f:mr}. The completeness mass limit we
  choose is that for the passive galaxies population, which
  guarantees our sample is also complete for the population of SF
  galaxies since they are intrinsically less massive than passive
  galaxies at a given magnitude.

\begin{figure}[ht]
 \centering
\includegraphics[width=1.\columnwidth]{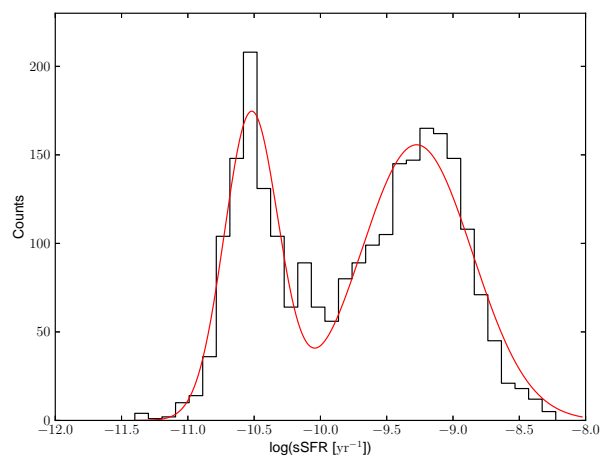}
\caption{Distribution of the sSFR for the total sample of
    cluster galaxies. The red curve represents the best-fit to this
    distribution with two Gaussians.}
\label{f:ssfr}
\end{figure}

In addition to $\mste$, among all the parameters provided by the
\texttt{MAGPHYS} procedure, we also consider the specific star
formation rate  (i.e., star formation rate per unit mass,
sSFR $\equiv$ SFR/$\mste$). The sSFR values are used to distinguish between
SF and passive galaxies. Even if we do not expect the
sSFR estimates from optical SED fitting to be very accurate, they are
sufficiently good to allow identification of the well-known bimodality
in the galaxy distribution (see Sect.~\ref{s:intro}). This
  can be better appreciated by looking at the sSFR distribution of
  cluster galaxies, shown in Fig.  \ref{f:ssfr}. This distribution is
  clearly bimodal.  Following \citet[][and references
  therein]{laralopez2010} we use the value sSFR$=10^{-10}$ yr$^{-1}$
to separate the populations of SF and passive
galaxies. This value also corresponds to a local minimum in the
  sSFR distribution.

In the public \texttt{MAGPHYS} library, there are many more dusty and
SF models than passive models (E. da Cunha, priv. comm.).  Whenever an
optical SED can be equally well fitted by a passive model and by a
dusty SF model, the median solution is biased in favor of dusty SF
models, since they occupy a larger area of the parameter space than
passive models.  It is therefore possible that some truly passive
  galaxies are classified as dusty SF galaxies. To estimate how
  serious this misclassification might be, we fit the sSFR
  distribution with two Gaussian distributions (see Fig.
  \ref{f:ssfr}). We make the hypothesis that misclassified passive
  galaxies lie in the high-sSFR tail of the Gaussian centered at low
  sSFR. The fraction of the area occupied by this Gaussian at
  sSFR$>10^{-10}$ yr$^{-1}$ is 0.5\%, and this is our estimate of the
  fraction of passive galaxies misclassified as SF. Similarly, one can
  estimate that the fraction of SF galaxies incorrectly classified as
  passive is 4\%. Given that these fractions are small, we consider
  our sSFR estimates sufficiently good to separate our sample into the
  two populations of passive and SF galaxies.

In order to check the reliability of our $\mste$ estimates, we make
use of the data from the UltraVista
survey\footnote{\texttt{http://home.strw.leidenuniv.nl/~ultravista/}}
(\citealt{mckracken2013}) which is an ultra-deep, near-infrared survey
with the VISTA survey telescope of the European Southern Observatory.
From the UltraVista public catalog we select only `USE = 1'
objects, i.e., objects classified as galaxies, with a K magnitude above
the detection limit of 23.9, and with uncontaminated and accurate
photometry (\citealt{muzzin2013a}).  We select only galaxies with
masses larger than our completeness limit ($\mathrm{10^{9.5}\,
  \Msun}$), and in the same photometric redshift range $\mathrm{0.38\,
  \leq\, z_{phot} \, \leq 0.50}$ used for our cluster membership
selection -- UltraVista $\mathrm{z_{phot}}$ have been obtained with
the EAZY code of \citet{brammer2008}.  To separate the UltraVista
sample into the passive and SF populations we use the separations
provided by \citet{muzzin2013b} in the UVJ diagram.

We compare the masses provided in the UltraVista data-base \citep[and
  obtained using the \texttt{FAST} SED-fitting code of][]{kriek2009}
with those we obtained applying \texttt{MAGPHYS} on the UltraVista
photometric catalog, using all of the available 30 bands, which cover 
the ultraviolet to mid-infrared, $\mathrm{24\, \mu m}$, spectral range. We find a good
agreement between the two $\mste$ estimates, apart from a median shift
$\mathrm{\Delta M \, =\, \log(M/\Msun)_{MAGPHYS}\, - \,
  \log(M/\Msun)_{FAST} \, =\, -0.07}$ independent from the galaxy type
and mass. This comparison suggests that the $\mste$ estimates are not
strongly dependent on the adopted SED-fitting algorithm, since
the mass difference is well below the typical uncertainty in
the individual $\mste$ estimates.

We then use the UltraVista dataset to check the effect of
using only the optical bands in the SED fitting. In fact, for our
analysis of the cluster SMF we can only use the optical SUBARU bands
(BVriz) over the whole cluster field. For this test, we apply
\texttt{MAGPHYS} to the selected UltraVISTA dataset once using all
available bands, and another time using only the five optical SUBARU
bands. The $\mste$ estimates obtained using optical bands only are
systematically higher than those obtained using all available
bands, particularly for the passive galaxies. The median value of
the shift, $\mathrm{\Delta M\, = \, log(M/\Msun)_{all\, bands}\, - \,
  log(M/\Msun)_{optical}}$, is -0.07 for SF galaxies and -0.23 for
passive galaxies. The shift for the SF galaxies is small, well below
the typical uncertainty in individual $\mste$ estimates.  On the other
hand, the shift in mass for the passive galaxies is not negligible.

The reason for the systematic shift in the $\mste$ estimates of
passive galaxies is probably related to the fact that \texttt{MAGPHYS}, when run on its public library, tends to favor dusty
  SF models rather than passive modes, when they cannot be
  distinguished based on the available data. We therefore run
\texttt{MAGPHYS} again only on the sample of passive galaxies,
this time using a library of templates heavily biased to fit old
stellar populations with very little star formation (kindly provided
by E. da Cunha). Using this library, we find that the shift between
the masses estimated using all UltraVISTA bands, and those estimated
using only optical bands is reduced to -0.13.  Since this is within
the typical uncertainty in individual $\mste$ estimates, we consider
the new mass estimates to be acceptable.

Using this new library of passive models, we then redetermine the
stellar masses of the passive galaxies identified in the cluster
M1206, by running \texttt{MAGPHYS} again.
 
%% \begin{figure}[ht]
%%  \centering
%% \includegraphics[width=1.\columnwidth]{confronto_magphys_opt_all.eps}
%% \caption{Difference between the \texttt{MAGPHYS} $\mste$ estimates
%%   obtained using the photometry from all available bands and those
%%   obtained using only optical band photometry, for the selected sample
%%   of UltraVista galaxies.  Blue diamonds and red dots indicate SF and,
%%   resp., passive galaxies.  Error bars are 1 $\sigma$.}
%% \label{f:all_opt}
%% \end{figure}

\subsection{Completeness and membership corrections}
\label{ss:compl}
To determine the cluster SMF, we need to apply two corrections to the
observed galaxy counts. One is the correction for the incompleteness
of the sample of galaxies with $\mathrm{z_{phot}}$, which also
contains all the galaxies in the spectroscopic sample.  The second
correction is to account for interlopers in the sample of
photometrically selected members (their presence is evident from
Fig.~\ref{f:zzp}, note the red dots with z very different from the cluster mean
z).

\begin{figure}[ht]
 \centering
\includegraphics[width=1.\columnwidth]{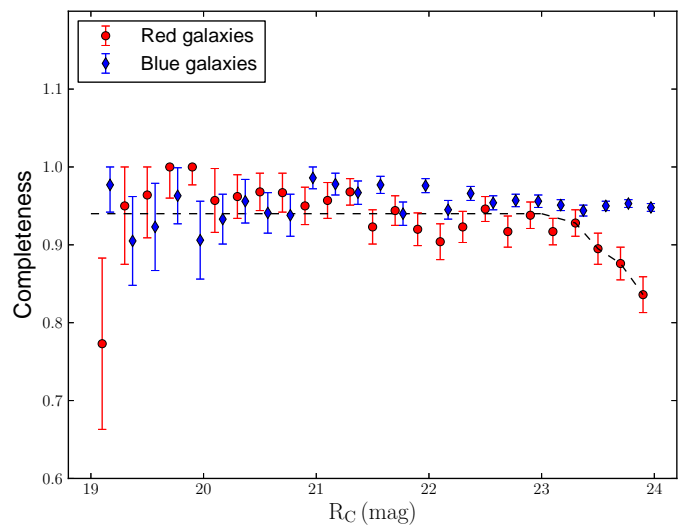}
\caption{Completeness of the $\mathrm{z_{phot}}$ sample as a function
  of the $\mathrm{R_C}$ magnitude, separately for red ($\mathrm{B\,
    -\, R_C} \geq 1.5$, red dots) and blue ($\mathrm{B\, -\, R_C} <
  1.5$, blue diamonds) galaxies. The dashed line represents the
  adopted completeness as a function of $\mathrm{R_C}$ for the sample
  of passive members, and, down to $\mathrm{R_C}=23$, for the
  sample of SF members also.  For $\mathrm{R_C}>23$, we adopt the
  same completeness used for brighter SF members.}
\label{f:compl}
\end{figure}
 
We estimate the completeness, C, of the sample of galaxies with
$\mathrm{z_{phot}}$ by measuring the ratio between the number of
galaxies with $\mathrm{z_{phot}}$, $\mathrm{N_{zp}}$, and the number
of galaxies in the $\mathrm{R_C}$ photometric sample, $\mathrm{N_p}$,
C $\equiv \mathrm{N_{zp}/N_p}$. In Fig. \ref{f:compl}, we show this
completeness in different magnitude bins, for red and blue
galaxies separately, where we use a color $\mathrm{B\, -\, R_C}=1.5$ to separate
the two samples. This value corresponds to the sSFR value
used to separate passive and SF members (see Sect.~\ref{ss:mass}),
and can therefore be used as a proxy for distinguishing these two
populations when sSFR estimates are not available. In fact, sufficient photometric information is not available for all of the
$\mathrm{N_p}$ galaxies to allow for a reliable sSFR estimate to be obtained from SED fitting.

Completeness is $>90$\% down to $\mathrm{R_C\, =}$ 23. In this
magnitude range the variation of C with $\mathrm{R_C}$ is negligible
and C is not significantly different for the red and blue samples. We
therefore adopt the value C= 0.94. In the magnitude range
$\mathrm{23\, \leq R_C\, \leq 24}$, we adopt the same C value for the
sample of SF galaxies, while for the passive galaxies we apply a
magnitude-dependent correction (using the values shown by the red dots
in Fig. \ref{f:compl}). We do not consider galaxies with
$\mathrm{R_C\, >}$ 24 in our analysis; on average in our sample this
magnitude limit corresponds to $\mste < \mathrm{10^{9.5}\, \Msun}$
(see Fig.~\ref{f:mr}).  Down to this limiting $\mste$ there are 1363
cluster members, of which 462 are spectroscopically confirmed
(i.e., $\sim 1/3$ of the total).  We define the correction factor for
incompleteness as $\mathrm{f_C}=\mathrm{1/C}$.

\begin{figure}[hŧ]
 \centering
\includegraphics[width=1.\columnwidth]{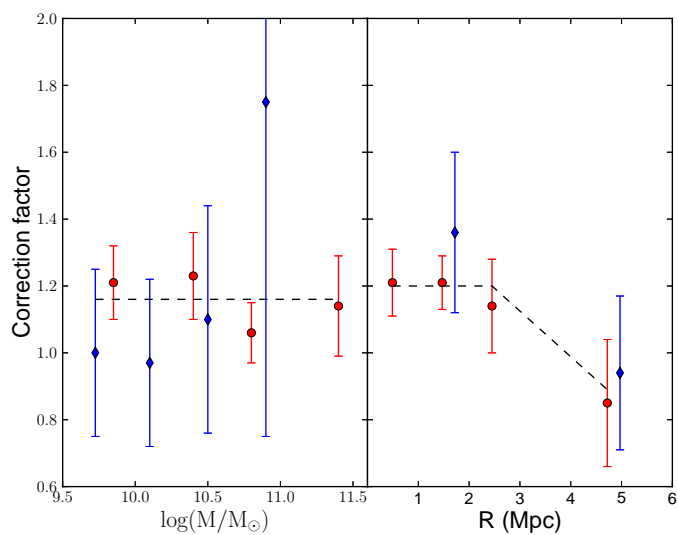}
\caption{The membership correction factor $\mathrm{f_{M}}$ (see text) as a
  function of $\mste$ (left panel) and clustercentric radius R (right
  panel). We adopt a correction factor independent of $\mste$ and
  dependent on R, the same for the samples of passive (red dots) and
  SF (blue diamonds) members.}
\label{f:fm}
\end{figure}

As for the membership correction of the $\mathrm{z_{phot}}$ sample, we
follow the approach of \cite{vanderburg2013}. We define the purity, P,
of the sample of photometric members as the ratio between the number
of photometric members that are also spectroscopic members, and the
number of photometric members with z, $\mathrm{P}=\mathrm{N_{pm\,\cap
    \, zm}}/\mathrm{N_{pm\, \cap \, z}}$. Since some real members are
excluded by the $\mathrm{z_{phot}}$ and color membership selection, we
need to define another completeness, given by the ratio between the
number of spectroscopic members that are also photometric members and
the number of spectroscopic members $\mathrm{C_M}=\mathrm{N_{pm\, \cap
    \, zm}}/\mathrm{N_{zm}}$. The membership correction
factor of the $\mathrm{z_{phot}}$ sample is then given by
$\mathrm{f_{M}}\equiv \mathrm{P/C_M}$. There is no significant dependence of $\mathrm{f_M}$ from the
galaxy $\mste$ (see Fig. \ref{f:fm}, left panel), but it does depend
mildly on projected clustercentric distance, R (see Fig. \ref{f:fm},
right panel).  This dependence is similar for passive and SF galaxies,
so we adopt the one evaluated for the passive sample also for the SF
sample. 

Note that the completeness correction factor $\mathrm{f_{C}}$ applies
to the full sample of photometric and spectroscopic members, since the
sample of galaxies with z is a subset of the sample of galaxies with
$\mathrm{z_{phot}}$, while the membership correction factor
$\mathrm{f_{M}}$ only applies to the sample of photometric members,
since the membership based on z is considered to be correct.

\section{The stellar mass function}
\label{s:smf}
We derive the cluster SMF by counting the number of cluster members
(defined in Sect.~\ref{s:data}) per bin of $\mste$, and correcting
these counts as described in Sect.~\ref{ss:compl}. The resulting
$\mste$ distribution is shown in Fig. \ref{f:smf} for all cluster
members, and also, separately, for passive and SF cluster members. We
estimate the errors in the galaxy counts with the bootstrap procedure
(Efron and Tibshirani 1986).

We fit these SMFs with a \citet{schechter1976} function
\begin{equation}
 \mathrm{\Phi (\log M) = \ln(10) \, \Phi^* \,
   \left(\frac{M}{M^*}\right)^{1+\alpha}
   \exp\left(-\frac{M}{M^*}\right) \, d(\log M),}
\label{eq:schec}
\end{equation}
where $\mathrm{\Phi^*}$ is the normalization, $\mathrm{\alpha}$ is the
low-mass end slope, and $\mathrm{M^*}$ corresponds to the exponential
cutoff of the SMF at high masses. The fits are performed down to the
mass limit $\mathrm{10^{9.5}\, M_{\odot}}$ (see Sect.~\ref{ss:compl}),
using the maximum likelihood technique \citep{MK86}. This
  technique has the advantage that no binning of the data is required.
  The normalization $\mathrm{\Phi^*}$ is not a free parameter, since
  it is constrained by the requirement that the integral of the
  fitting function over the mass range covered by observations equals
  the number of galaxies in the sample. Of course, this number must be
  corrected for completeness and membership contamination.
  Therefore, in the maximum likelihood fitting procedure, the
  product of the completeness and membership correction factors, 
  $\mathrm{f_C \cdot f_M}$, are used as weights for the individual
  values of $\mste$.  Therefore, there are only two free parameters in the fit, $\mathrm{\alpha}$ and $\mathrm{M^*}$, except
    when we fit the data with a double Schechter function (in
    Sect.~\ref{ss:type}),
\begin{eqnarray}
   \mathrm{\Phi (\log M) = \ln(10) \, \Phi^* \, \times} & \nonumber \\
   \mathrm{\left[\left(\frac{M}{M^*}\right)^{1+\alpha}
   \exp\left(-\frac{M}{M^*}\right) + f_2
   \left(\frac{M}{M_2^*}\right)^{1+\alpha_2}
   \exp\left(-\frac{M}{M_2^*}\right) \right] \, d(\log M).} &
\label{eq:double}
\end{eqnarray}
In this case there are three additional free parameters,
  $\mathrm{\alpha_2}$ and $\mathrm{M^*_2}$, and the ratio between the
  normalizations of the two Schechter functions,
  $\mathrm{f_2=\Phi^*_2/\Phi^*}$.

In the fits, in addition to the statistical errors, we also take into
consideration the errors on the stellar mass estimates.  These
are evaluated by performing 100 Monte-Carlo simulations in which the
mass of each galaxy is extracted randomly from a Gaussian distribution
centered on the best-fit mass value, with a standard deviation equal
to the error on the mass estimate.  The errors on the stellar mass
estimates provide only a minor contribution to the uncertainties on
the best-fit Schechter function parameters, which are dominated by the
statistical errors on the number counts.

We do not take the errors on the photometric
  completeness into account (see Fig.~\ref{f:compl} in Sect.~\ref{ss:compl}) since
  they are small. They would affect mostly the normalization of the
  SMF, while in most of our analyses we are only interested in
  comparing the shapes of different SMFs.  We only care about the
  normalization of the SMF when comparing the passive and SF samples
  (Sect.~\ref{ss:type}) and when comparing the SMF in the central
  cluster region to the mass in the intracluster light (ICL, see
  Sect.~\ref{s:disc}). Also, in these cases, we estimate that the
  errors on the completeness can be neglected without a significant
  impact on our results.

The errors on the membership correction factor $\mathrm{f_M}$ (see
Fig.~\ref{f:fm} in Sect.~\ref{ss:compl}) are significantly larger than
those on the completeness.  We estimate their effect on the stellar
mass function in the following way. First, we consider the
effect of adopting different values of $\mathrm{f_M}$ for red (passive)
and blue (SF) galaxies, given by their different means, rather than
adopting the same value for both populations. Second, in the regions
where $\mathrm{f_M}$ deviate from a constant, i.e., at
$\mathrm{R\,>\,r_{200}}$, we consider the effect of adopting
the two extreme values of $\mathrm{f_M}$ given by $\mathrm{f_M\, \pm\,
  \sigma_{f_{M}}}$, where $\mathrm{\sigma_{f_{M}}}$ is the error in
our estimate of $\mathrm{f_M}$ at $\mathrm{R\,>\,r_{200}}$. We find
that all the results of the analyses presented in the following
sections do not change significantly when changing the membership
correction factors as described above.  We therefore conclude that
the uncertainties on $\mathrm{f_M}$ do not have a significant impact on
our results.  For the sake of clarity, in the following sections we
only present the results based on our best estimates of
$\mathrm{f_M}$, i.e., those given in Sect.~\ref{ss:compl}.

We assess the statistical significance of the difference between any two
SMFs both parametrically, by comparing the best-fit parameters of the
Schechter function, and non parametrically, via a Kolmogorov-Smirnov
(K-S) test (e.g., \citealt{press2013}). We require a minimum of ten
objects in a sample for a meaningful comparison. 

The K-S test compares the cumulative distributions and therefore it is
only sensitive to differences in the shapes of the
distributions, not in their normalizations. However, in most of
  our analysis we are not interested in the normalization of the SMF, rather in its shape.  There are only two
  points in our analysis where the normalization of the SMF is
  important. One is in the comparison of the passive and SF
  populations (see Sect.~\ref{ss:type}), since different relative
  normalizations affect the mass value at which the two SMFs cross
  each other -- a useful parameter to constrain theoretical models
  (see Sect.~\ref{s:disc}). Another point is the estimate of the mass
  that could have been stripped from galaxies and gone into the mass
  of the ICL (see Sect.~\ref{s:disc}). In other parts of our
  analysis, differences in the SMF normalization just reflect rather
obvious dependencies of the number densities of galaxies (of different
types) on the environments where they are located, e.g., the cluster
is denser than the field by definition, and this over-density is
higher among the population of passive galaxies by virtue of the
well-known morphology-density relation (\citealt{dressler1980}). 
  For the comparison of the SMFs of a given cluster galaxy population in
  different environments, the K-S test is particularly appropriate.
For the same reason, to highlight differences in the SMFs, we
only compare the shape parameters of the Schechter function best-fits,
$\mathrm{\alpha}$ and $\mathrm{M^*}$, and not the normalization
parameter $\mathrm{\Phi^*}$.

\begin{table*}[ht]
\centering
\caption{Best-fit Schechter function parameters}
\label{t:sbf1}
%\small
% \begin{tabular}{p{4cm} p{2cm} p{2cm}}
\begin{tabular}{lcccccc}
 \hline
Galaxy type & $\mathrm{\Phi^*}$ & $\mathrm{\alpha}$ & $\mathrm{log(M^*/\msun)}$ &
$\mathrm{f_{2}}$ & $\mathrm{\alpha_{2}}$ & $\mathrm{log(M^*_2/\msun)}$\\
\hline
Passive  & 654 & -0.38 $\mathrm{\pm}$ 0.06 & 10.96 $\mathrm{\pm}$ 0.04 & -- & -- & --\\
SF  & 156 & -1.22 $\mathrm{\pm}$ 0.10 & 10.68 $\mathrm{\pm}$ 0.09 & -- & -- & --\\
All  & 751 & -0.39 $\mathrm{\pm}$ 0.18 & 10.94 $\mathrm{\pm}$ 0.16 & 0.65 $\mathrm{\pm}$ 0.20 & -0.51 $\mathrm{\pm}$ 0.33 & 9.93 $\mathrm{\pm}$ 0.35 \\
All  & 541 & -0.85  $\mathrm{\pm}$ 0.04 & 11.09 $\mathrm{\pm}$ 0.04 & -- & -- & --\\
\hline
\end{tabular}
\tablefoot{$\mathrm{\Phi^*}$ 
is not a free parameter in the
fitting procedures, hence we do not provide error bars on its values.}
\end{table*}

The parametric comparison naturally takes the
completeness and membership corrections applied to the number
counts into account. These corrections are also taken into account in the K-S
tests, since we use the correction factors as weights in the evaluation of
the cumulative distributions whose maximum difference is used in the
test to evaluate the statistical significance of the null hypothesis.

\subsection{Different galaxy types}
\label{ss:type}
\begin{figure}[ht]
\centering \includegraphics[width=1.\linewidth]{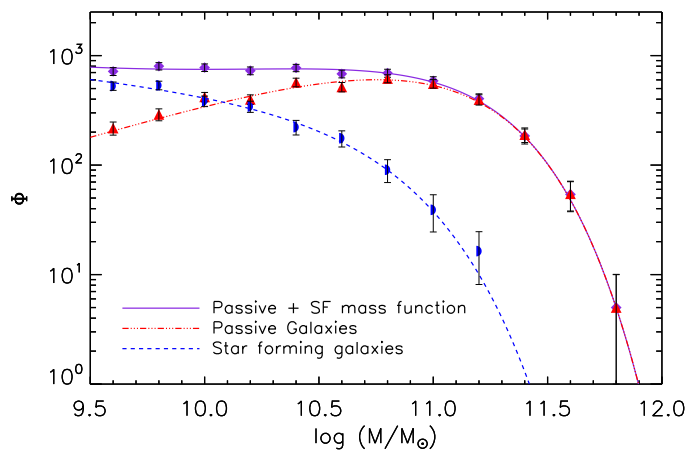}
  \caption{SMF for passive and SF cluster members (red
      triangles and blue demi-circles, respectively) and their
      best-fit Schechter functions (red triple-dot-dashed and blue
      dashed lines). The sum of the two SMFs is shown as
      a solid violet line.  Violet diamonds are the counts obtained
      by considering all cluster members.  The points represent
    counts in bins of 0.2 dex in $\mste$ divided for the bin
      size, and the counts have been corrected for completeness and
    membership. The (1~$\sigma$) errors on the counts have been
    estimated via the bootstrap resampling procedure.  }
  \label{f:smf}
  \end{figure}

\begin{figure}
 \includegraphics[width=1\linewidth]{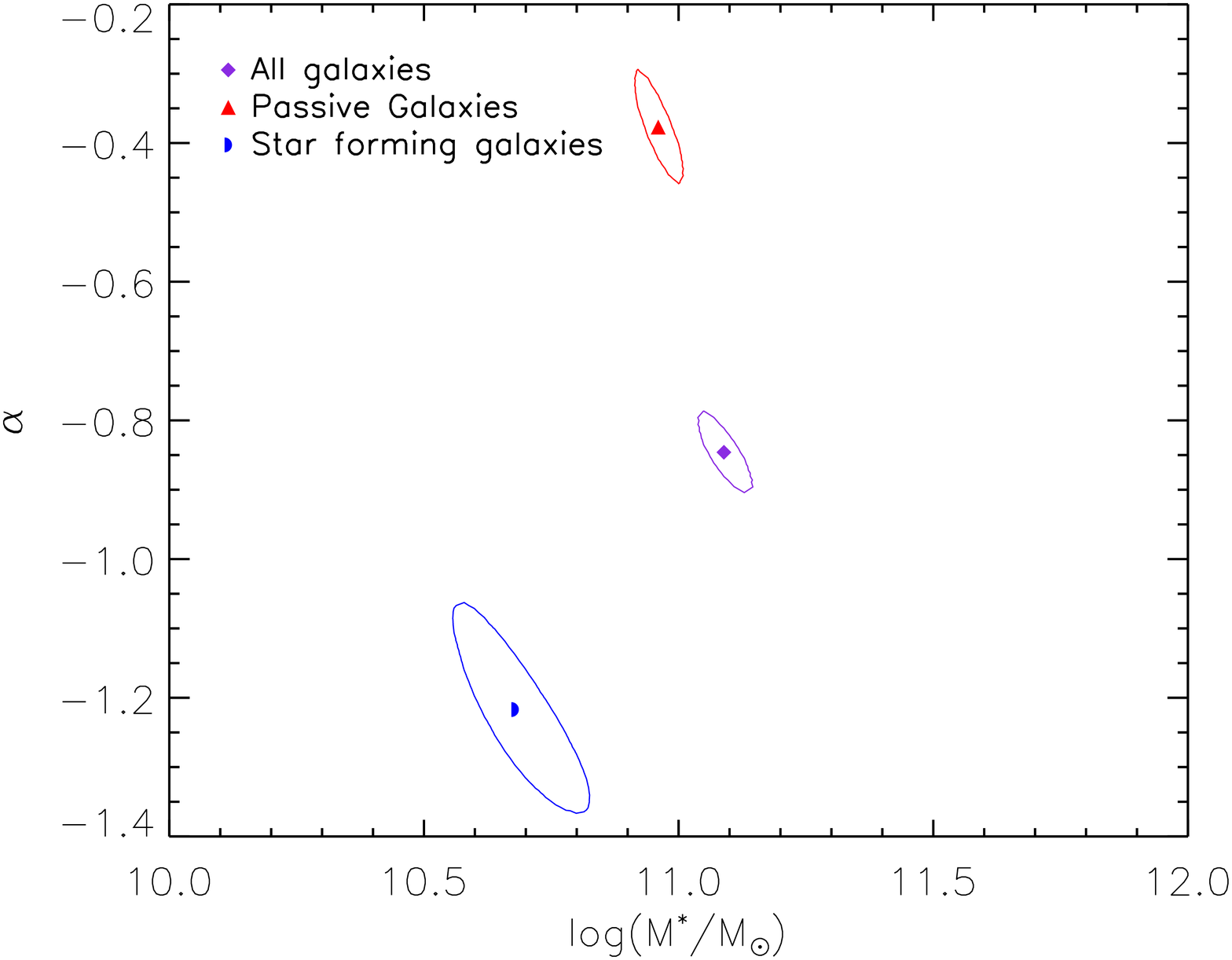}
 \caption{Best-fit Schechter parameters $\mathrm{M^*}$ and
   $\mathrm{\alpha}$ and 1~$\sigma$ likelihood contours.}
  \label{f:sbf}
  \end{figure}

In Fig. \ref{f:smf}, we show the SMF of the passive, and
  separately, the SF galaxy populations along with their
best-fitting Schechter functions\footnote{ In this and the following
    figures, the data are binned only for the sake of displaying the
    results of the fits. No binning of the data is required in the
    fitting procedure.}. The best-fit $\mathrm{\alpha}$ and
$\mathrm{M^*}$ parameters and their 1~$\sigma$ uncertainties, 
  obtained by marginalizing over the other free parameter, are shown
in Fig.  \ref{f:sbf} and listed in Table \ref{t:sbf1}.  When split
into the two cluster populations of passive and SF galaxies, the SMF
displays a strong, statistically significant dependence on galaxy
type.  In particular, the SMF of SF galaxies is increasing at the
low-mass end, while the SMF of passive galaxies is decreasing.  This
difference is also confirmed by the K-S test, which gives a very low
probability to the null hypothesis that the $\mste$ distribution of SF
and passive galaxies are drawn from the same population (see
Table~\ref{t:ks}).

This type-dependence of the SMF is not only valid in general for the
whole cluster, but also in different cluster regions, identified by
their clustercentric distance or by their local galaxy number density
in Sect.~\ref{sss:radial} and \ref{sss:dens} (see Table~\ref{t:ks}).

In Fig. \ref{f:smf}, we also show the sum of the two Schechter
  functions that describe the SMFs of passive and SF galaxies. We have also performed a fit of a single Schechter function to the SMF of all cluster galaxies together; the best-fit parameters for this function are given in Table \ref{t:sbf1}. According to the likelihood-ratio
  test \citep{Meyer+75}, the sum of the  two Schechter
    functions provides a significantly better fit than the single Schechter function (with a probability of $>0.999$), after taking
  into account the difference in the number of free parameters (four
  vs. two).

We also fit the SMF of all galaxies with a double Schechter
(eq.~\ref{eq:double}) and five free parameters, namely the
$\mathrm{\alpha}$ and $\mathrm{M^*}$ of the two Schechter functions
and their relative normalization. The best-fit parameters are listed in
Table \ref{t:sbf1} along with their marginalized errors. The best-fit parameters of one of the two Schechter
functions are very similar to those of the Schechter function that
provides the best-fit to the SMF of passive galaxies. On the other
hand, the best-fit parameters of the other Schechter function are very
different from those of the Schechter function that provides the
best-fit to the SMF of SF galaxies. This means that while the best-fit
with a double Schechter is optimal from a statistical point of view,
it fails to correctly describe one of the two components of the
cluster galaxy sample, that of SF galaxies. This is probably because of 
the fact that the sample of cluster galaxies is largely dominated by
passive galaxies over most of the mass range covered by our analysis,
and so it is difficult to correctly identify the minority component,
that of SF galaxies. As a matter of fact, the uncertainties on the
best-fit parameters of the double Schechter function are rather
large.

\begin{table}[ht]
\centering
\caption{Best-fit Schechter function parameters for different environments}
\label{t:sbf}
%\small
% \begin{tabular}{p{4cm} p{2cm} p{2cm}}
\begin{tabular}{llcc}
 \hline
Galaxy type & Environment & $\mathrm{\alpha}$ & $\mathrm{log(M^*/\msun)}$\\
\hline
\\
SF & $\mathrm{R > r_{200}}$ & -1.07 $\mathrm{\pm}$ 0.12 & 10.54 $\mathrm{\pm}$ 0.09\\
SF & $\mathrm{R \leq r_{200}}$ & -1.52 $\mathrm{\pm}$ 0.17 & 11.16 $\mathrm{\pm}$ 0.37\\
\\
Passive & $\mathrm{R > r_{200}}$ & -0.43 $\mathrm{\pm}$ 0.09 & 10.99 $\mathrm{\pm}$ 0.05\\ 
Passive & $\mathrm{R \leq r_{200}}$ & -0.40 $\mathrm{\pm}$ 0.08 & 11.00 $\mathrm{\pm}$ 0.05\\
\\
Passive & Region 1 & -0.15 $\mathrm{\pm}$ 0.15 & 10.92 $\mathrm{\pm}$
0.08\\
Passive & Region 2 & -0.54 $\mathrm{\pm}$ 0.14 & 10.97 $\mathrm{\pm}$
0.10\\
Passive & Region 3 & -0.44 $\mathrm{\pm}$ 0.15 & 10.99 $\mathrm{\pm}$
0.10\\
Passive & Region 4 & -0.57 $\mathrm{\pm}$ 0.14 & 11.09 $\mathrm{\pm}$ 0.11\\
\\
Passive & Region (a) & -0.13 $\mathrm{\pm}$ 0.16 & 10.93 $\mathrm{\pm}$
0.08\\
Passive & Region (b) & -0.55 $\mathrm{\pm}$ 0.16 & 11.00 $\mathrm{\pm}$
0.12\\
Passive & Region (c) & -0.41 $\mathrm{\pm}$ 0.11 & 10.95 $\mathrm{\pm}$
0.07\\
Passive & Region (d) & -0.35 $\mathrm{\pm}$ 0.09 & 10.96 $\mathrm{\pm}$ 0.06\\
\hline
\end{tabular}
%% \tablefoot{The results for the `Field' environment are based on data from the
%% UltraVista survey \citealt{mckracken2013}.}
\end{table}

\begin{table}[ht]
\centering
\caption{Results of the K-S tests}
\label{t:ks}
\begin{tabular}{lcr}
 \hline
Compared samples & N1, N2 & Prob. (\%) \\
 \hline
\\
\multicolumn{3}{c}{Type dependence}\\
\\
Passive vs. SF in the cluster  & 846, 517 & $<0.01$ \\
Passive vs. SF in Region 2 &  120, \phantom{0}21 & $<0.01$\\
Passive vs. SF in Region 3 &  120, \phantom{0}31 & $<0.01$\\
Passive vs. SF in Region 4 &  102, \phantom{0}20 & $<0.01$\\
Passive vs. SF in Region (b) & \phantom{0}96, \phantom{0}33 & 0.4\\
Passive vs. SF in Region (c) & 199, \phantom{0}54 & $<0.01$\\
Passive vs. SF in Region (d) & 328, 420 & $<0.01$\\
\\
\multicolumn{3}{c}{Environment dependence - SF galaxies}\\
\\
SF within and outside $\mathrm{r_{200}}$ & \phantom{0}78, 439 & $>10$\\
SF in Regions 2 and 3 & \phantom{0}31, \phantom{0}31  & $>10$\\
SF in Regions 3 and 4 & \phantom{0}31, \phantom{0}20  & $>10$\\
SF in Regions (b) and (c) & \phantom{0}33, \phantom{0}54  & $>10$\\
SF in Regions (c) and (d) & \phantom{0}54, 422 & $>10$\\
\\
\multicolumn{3}{c}{Environment dependence - passive galaxies}\\
\\
Passive within and outside $\mathrm{r_{200}}$ & 438, 408 & $>10$\\
Passive in Regions 1 and 2 & 120, 120 & 0.8\\
Passive in Regions 2 and 3 & 120, 102 & $>10$\\
Passive in Regions 3 and 4 & 102, \phantom{0}96 & $>10$\\
Passive in Regions (a) and (b) & 100, \phantom{0}83 & 0.4\\
Passive in Regions (b) and (c) & \phantom{0}83, 199 & $>10$\\
Passive in Regions (c) and (d) & 199, 328 & 2\\
\\
\hline
\end{tabular}
\tablefoot{N1 and N2 are the number of galaxies in the two compared
  samples. The listed probabilities (Prob., in \%) are for the null
  hypothesis that two $\mste$ distributions are drawn from the same
  parent population. Probabilities $>10$\% indicate statistically
  indistinguishable distributions. Only the distributions of samples
  with at least ten objects have been considered.}
\end{table}

\subsection{Different environments}
\label{ss:env}
To search for possible environmental dependences of the SMF, we
separate passive and SF galaxies in this analysis, to disentangle
possible type-specific environmental dependences of the SMF from the
well-known environmental dependence of the galaxy population
(\citealt{dressler1980}; \citealt{baldry2008} and references therein).

We adopt two definitions of `environment', one based on the distance
from the cluster center, and another based on the local number density
of cluster members. Of course, these definitions are not entirely
independent, given the correlation between local density and radial
distance \citep[e.g.][]{Whitmore+93}.  Using these two definitions we
define nine cluster regions, five at different distances from
the cluster center, (four within $\mathrm{r_{200}}$, labeled 1 to
  4, and another one at $\mathrm{R>r_{200}}$, see
Sect.~\ref{sss:radial}) and another four at different local densities
(labeled (a) to (d), see Sect.~\ref{sss:dens}).

Our first definition of environment assumes circular symmetry,
  but the cluster is significantly elongated in the plane of the sky
  \citep{umetsu2012}.  However, this assumption is dropped in our
  other definition, as the regions (a) to (c) are elongated in the
  direction traced by the galaxy distribution (see Fig.~\ref{f:dens}),
  which is similar to the elongation direction of the brightest
  cluster galaxy (BCG) and of the total mass of the cluster as inferred from
  a weak lensing analysis by \citet[][see their Figs.~1 and
    11]{umetsu2012}. As we show below, our results are essentially
  independent on which definition of environment we adopt, hence the
  assumption of circular symmetry does not seem to be critical.

 \begin{figure}[ht]
 \includegraphics[width=1.\linewidth]{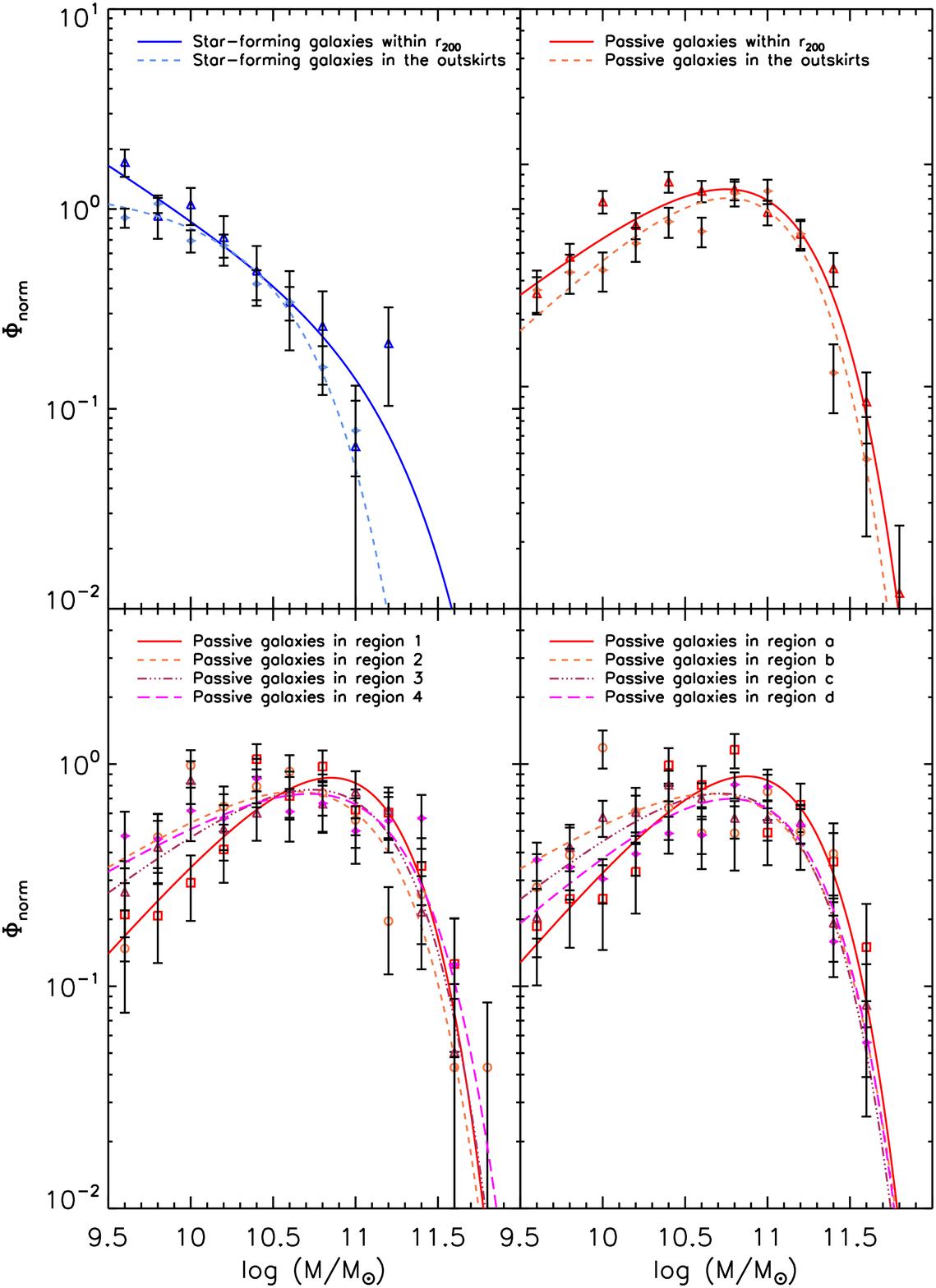}
  \caption{SMFs of SF and passive galaxies in different cluster
    regions and in the field.  Upper left (resp. right) panel: SMFs of
    SF (resp. passive) cluster galaxies beyond and within
    $\mathrm{r_{200}}$. Bottom left panel: SMFs of passive cluster
    galaxies in four different regions, defined by their distances
    from the cluster center (see text). Bottom right panel: SMFs of
    passive cluster galaxies in four different regions, defined by
    their local number densities. SMFs are normalized to the total
    number of galaxies contained in the respective samples.}
  \label{f:smfs}
  \end{figure}

 \begin{figure}[ht]
\includegraphics[width=1.\linewidth]{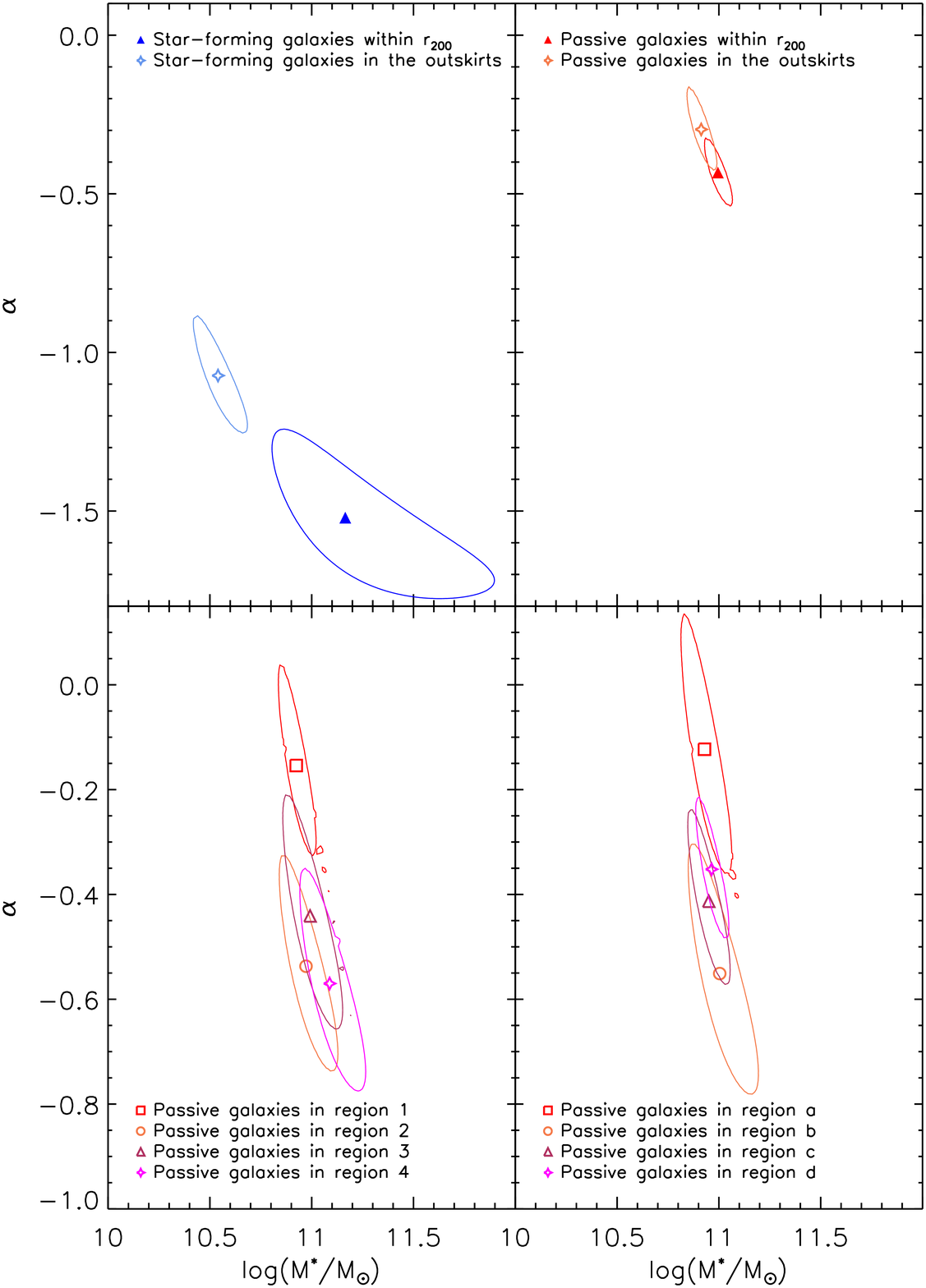}
  \caption{Best-fit Schechter parameters $\mathrm{M^*}$ and
    $\mathrm{\alpha}$ and 1~$\sigma$ likelihood contours, after
    marginalizing over the $\mathrm{\Phi^*}$ parameter for the SMFs of
    SF and passive galaxies in different cluster regions and in the
    field. The panels correspond one-to-one to those of
    Fig.~\ref{f:smfs}.}
  \label{f:sbfs}
  \end{figure}

\subsubsection{Clustercentric radial dependence}
\label{sss:radial}
We consider here the clustercentric distance as a definition of
`environment'. The cluster center is identified with the position of
the BCG \citep[see Table~\ref{t:props} and][]{biviano2013}.

We first consider the SMFs of passive and, separately, the SF cluster
members, within and outside the virial radius,
$\mathrm{r_{200}}$. These are shown in Fig.~\ref{f:smfs} (upper
panels), and their best-fit Schechter function parameters are listed
in Table~\ref{t:sbf} and shown in Fig.~\ref{f:sbfs} (upper 
panels).  The SMFs of cluster members within and outside the
virial radius are not significantly different, neither for the
passive nor for the SF galaxies. This is confirmed by the K-S test
(see Table~\ref{t:ks}).

  \begin{figure}[ht]
\centering
 \includegraphics[width=\linewidth]{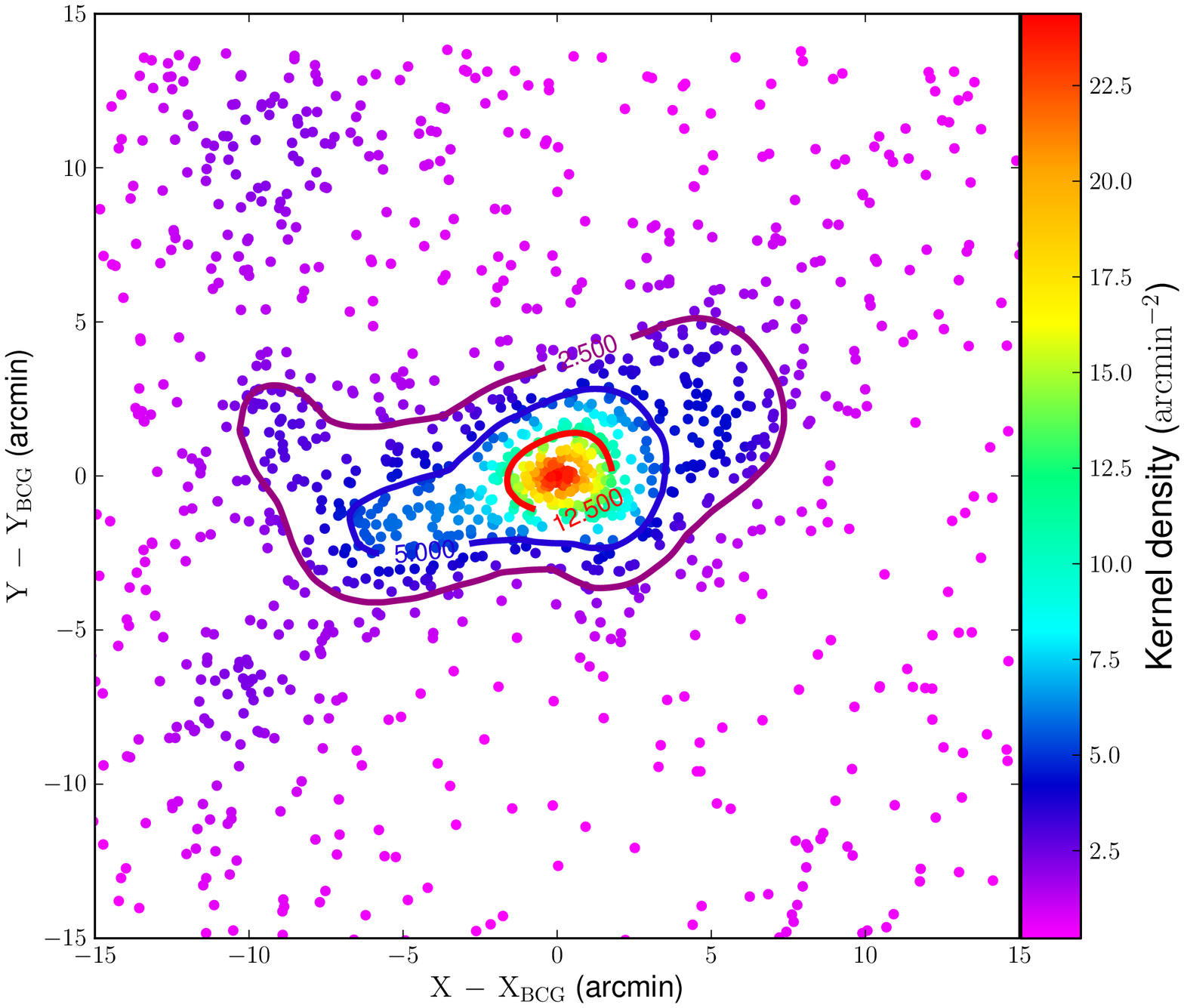}
  \caption{Spatial distribution of cluster members. The local number
    density is color coded as indicated by the bar at the right side
    of the plot. Coordinates are in arcmin with respect to the
    position of the BCG, see Table~\ref{t:props}. Galaxies
      belonging to Regions (a), (b), (c), which are defined in
      Sect.~\ref{sss:dens}, are those inside the red, blue, and purple
      solid lines, respectively. Galaxies belonging to Region (d) are
      the outer points.}
  \label{f:dens}
  \end{figure}

We then determine the SMF in four different regions within
$\mathrm{r_{200}}$, namely (see also Fig.~\ref{f:dens}):
\begin{enumerate}
 \item $\mathrm{R/r_{200}\, \le \, 0.25}$.
 \item $\mathrm{0.25\, < \, R/r_{200}\, \le\, 0.5}$.
 \item $\mathrm{0.5\, < \, R/r_{200}\, \le\, 0.75}$.
 \item $\mathrm{0.75\, < \, R/r_{200}\, \le\, 1}$.
\end{enumerate}
The number of SF galaxies is not large enough to allow for Schecter function fits in all these regions, however, in some cases we
have enough galaxies in the subsamples to allow for K-S test comparisons
of the $\mste$ distributions. On the other hand, we have a
sufficiently large number of passive galaxies to allow for meaningful
Schechter function fits in all the four regions. The SMFs for the
passive galaxies in the four different regions are shown in
Fig. \ref{f:smfs} (bottom left panel), along with their best-fitting
Schechter functions. The best-fit parameters are listed in
Table~\ref{t:sbf} and shown in Fig.~\ref{f:sbfs} (bottom left panel).

To highlight a possible radial dependence of the cluster SMF we
compare the $\mste$ distributions of cluster members in adjacent
regions, using the K-S test, separately for SF and passive galaxies,
whenever there are at least ten galaxies in each of the subsamples.
The $\mste$ distributions of the SF cluster galaxies in the different
regions are not statistically different (see Table~\ref{t:ks}). On the
other hand, the K-S tests indicate a significant difference of the
$\mste$ distributions of passive cluster members in Region 1 (the
innermost one) and the adjacent Region 2.  For no other adjacent
regions does the K-S test highlight a significant difference from the
SMFs of passive galaxies.

 \begin{figure}[ht]
\includegraphics[width=1.\linewidth]{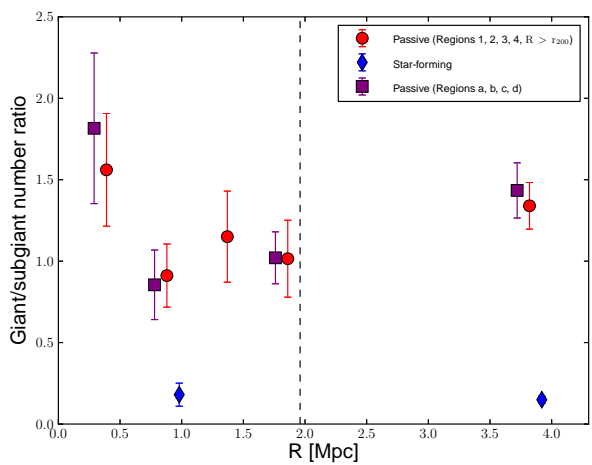}
  \caption{The number ratio of giant galaxies ($\mathrm{\log
      \mste/\msun \geq 10.5}$) and subgiant galaxies ($\mathrm{\log
      \mste/\msun < 10.5}$), GSNR, for different samples of passive
    galaxies (red dots: Regions 1--4 and $\mathrm{R>r_{200}}$;
    magenta squares: Regions (a)--(d)) and SF galaxies (blue diamonds:
    within and outside the virial radius). For Regions (a)--(d)
    the point abscissae are set at the average clustercentric radii of
    the galaxies in the subsamples selected on the base of local
    density. The vertical dashed line indicates the location of
    $\mathrm{r_{200}}$.  }
  \label{f:ratio}
  \end{figure}

From Fig.~\ref{f:sbfs} (bottom left panel) and Table~\ref{t:sbf} one
can see that the difference of the SMFs in Regions 1 and 2 is
reflected in a difference in the values of the best-fit Schechter
parameter $\alpha$. From Fig.~\ref{f:smfs} (bottom left panel) we can
indeed see that the SMF of passive galaxies in Region 1 is
characterized by a low-mass end drop that is more rapid than
for the SMFs in other Regions.

From Fig.~\ref{f:sbfs}, one can notice that the SMFs of passive
galaxies in Region 1 intersects those of passive galaxies in the other
regions at $\mathrm{\log(\mste/\msun)\sim10.5}$. Since these SMFs
  are normalized by the total number of galaxies in their respective
  samples, this does not mean that in Region 1 there are more galaxies
  with $\mathrm{\log(\mste/\msun)\sim10.5}$ than in other
  regions. This mass value only indicates where the relative ratio of
  the number of galaxies more massive and less massive
  than a given value is maximally different for the SMF in Region 1
  and in the other regions.  We therefore use this value to
separate `giant' from `subgiant' galaxies and plot the giant/subgiant
number ratio (GSNR hereafter) as a function of radial distance from
the cluster center in Fig.~\ref{f:ratio}. Note that we use the
correction factors defined in Sect.~\ref{ss:compl} as weights to
compute the GSNR.  The GSNR of passive galaxies decreases rapidly from
the center (Region 1) to $\mathrm{R \sim 0.8}$ Mpc, then gently
increases again toward the cluster outskirts ($\mathrm{R > 3.5}$ Mpc)
but without reaching the central value again. The GSNR of SF galaxies
does not seem to depend on radius and is systematically below that of
passive galaxies at all radii.

\subsubsection{Density dependence}
\label{sss:dens}
As an alternative definition of `environment', we consider here the
local number density of cluster members. This density is defined
  by smoothing the projected distribution of galaxies with a
  two-dimensional Gaussian filter in an iterative way.  Initial
  estimates of the densities are obtained by using a fixed `optimal'
  (in the sense of \citealt{silverman1986}) characteristic width for
  the Gaussian filter.  In the second iteration, the characteristic
  width of the Gaussian filter is locally modified by inversely
  scaling the `optimal' width with the square root of the initial
  density estimates. In other words, we adopt an adaptive-kernel
  filtering of the galaxy spatial distribution, where the kernel is
  adapted in such a way as to be narrower where the density is
  higher.

The distribution of cluster
members is shown in Fig. \ref{f:dens}, where symbols are colored
according to the local galaxy density. We then define four regions of
different mean projected density $\Sigma$ (in units of arcmin$^{-2}$):
\begin{enumerate}[(a)]
 \item $\mathrm{\Sigma >\, 12.5}$.
 \item $\mathrm{5 <\, \Sigma\, \le\, 12.5}$.
 \item $\mathrm{2.5 <\, \Sigma\, \le\, 5}$.
 \item $\mathrm{\Sigma\, \le\, 2.5}$.
\end{enumerate}
From Fig. \ref{f:dens} one can note that Regions (a) and (b)
approximately correspond to Regions 1 and 2 (defined in
Sect.~\ref{sss:radial}), while Region (c) corresponds roughly to
Regions 3 and 4 with an extension beyond the virial radius. One
obvious difference is that the regions defined by the value of
$\mathrm{\Sigma}$ are more elongated than those defined by radius.

Since there are not enough SF galaxies to allow for meaningful Schechter
fits to be performed in Regions (a) to (d), in Fig.~\ref{f:smfs}
(bottom right panel) we only show the passive galaxy SMFs and their
Schechter best fits. The best-fit parameters are listed in
Table~\ref{t:sbf} and shown in Fig.~\ref{f:sbfs} (bottom right panel).

The K-S tests indicate that the SMF of SF galaxies is independent of
local density (see Table~\ref{t:ks}). On the contrary, the SMF of
passive galaxies does depend on local density. In fact, the K-S tests
performed between $\mste$ distributions in adjacent regions indicate a
significant difference between Regions (a) and (b) (see
Table~\ref{t:ks}). This difference is caused by the more rapid drop at
the low-mass end of the SMF in Region (a) compared to the SMFs of
other regions (Fig.~\ref{f:sbfs}, bottom right panel) and is
reflected in a different value of the best-fitting parameter $\alpha$
(see Table~\ref{t:sbf}).

In Fig.~\ref{f:ratio}, we can see that the radial trend of the GSNR
of passive galaxies found in regions 1--4 is confirmed when considering
regions (a)--(d).

\section{The stellar mass density profile}
\label{s:smfrac}
Using the sample of 1363 cluster members with $\mste \geq
\mathrm{10^{9.5}\, M_{\odot}}$, we determine the radial profiles of
number and stellar mass density of our cluster, N(R) and
$\mathrm{\Sigma_{\star}(R)}$, respectively. We fit these profiles in
the region $0.05 < \mathrm{R/r_{200}} \leq 1$ (i.e., excluding the
BCG) with a projected NFW (pNFW) model \citep{NFW97,Bartelmann96}
using a weighted maximum likelihood fitting technique. For the
determination of N(R), we use as weights those already used for the
construction of the SMF, i.e., the product $\mathrm{f_C \cdot f_M}$
(see Sect.~\ref{ss:compl}).  For the determination of
$\mathrm{\Sigma_{\star}(R)}$, we use the same weights multiplied by
the galaxy stellar masses, $\mathrm{f_C \cdot f_M \cdot \mste}$. In
Table~\ref{t:rs}, we list the values of the scale radii,
$\mathrm{r_s}$, of the best-fit models. Note that our best-fit value
for the $\mathrm{r_s}$ of N(R) is consistent with that estimated by
\citep{biviano2013} on a slightly different sample.  We find that
$\mathrm{\Sigma_{\star}(R)}$ is significantly more concentrated than
N(R).

\begin{figure}[ht]
\includegraphics[width=1.\linewidth]{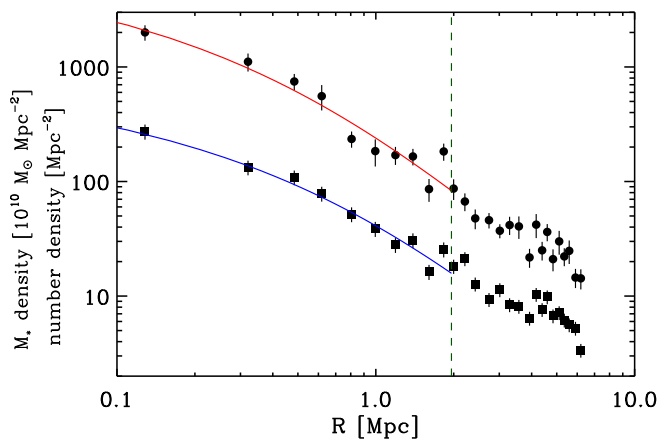}
  \caption{The stellar mass density profile (dots) and the number
    density profile (squares) and their best-fit projected NFW models
    (red and blue curves). The 1 $\sigma$ errors are shown,
    and evaluated using a bootstrap procedure. Both densities are space
    densities.  The vertical dashed green line indicates the location
    of $\mathrm{r_{200}}$.}
  \label{f:profs}
  \end{figure}

The two profiles and their best-fit models are shown in
Fig.~\ref{f:profs}. The error bars in the figure have been estimated
via a bootstrap procedure.  The pNFW model provides a good fit to the
number density profile (reduced $\chi^2=1.4$), and a slightly worse
fit to the stellar mass density profile (reduced $\chi^2=2.2$).

\begin{table}[ht]
\centering
\caption{The NFW scale radii of the density profiles}
\label{t:rs}
\begin{tabular}{ll}
 \hline
Profile & $\mathrm{r_s}$ (Mpc) \\
 \hline
\\
Galaxy number density & $0.80_{-0.14}^{+0.05}$ \\
\\
Stellar mass density & $0.52_{-0.07}^{+0.06}$ \\
\\
Total mass density & $0.34_{- 0.06}^{+0.06}$ \\
\\
\hline
\end{tabular}
\tablefoot{The scale radius of the total density profile is adopted from
  \citet{umetsu2012} \citep[see also Table 3 in][]{biviano2013}.}
\end{table}

 \begin{figure}[ht]
\includegraphics[width=1.\linewidth]{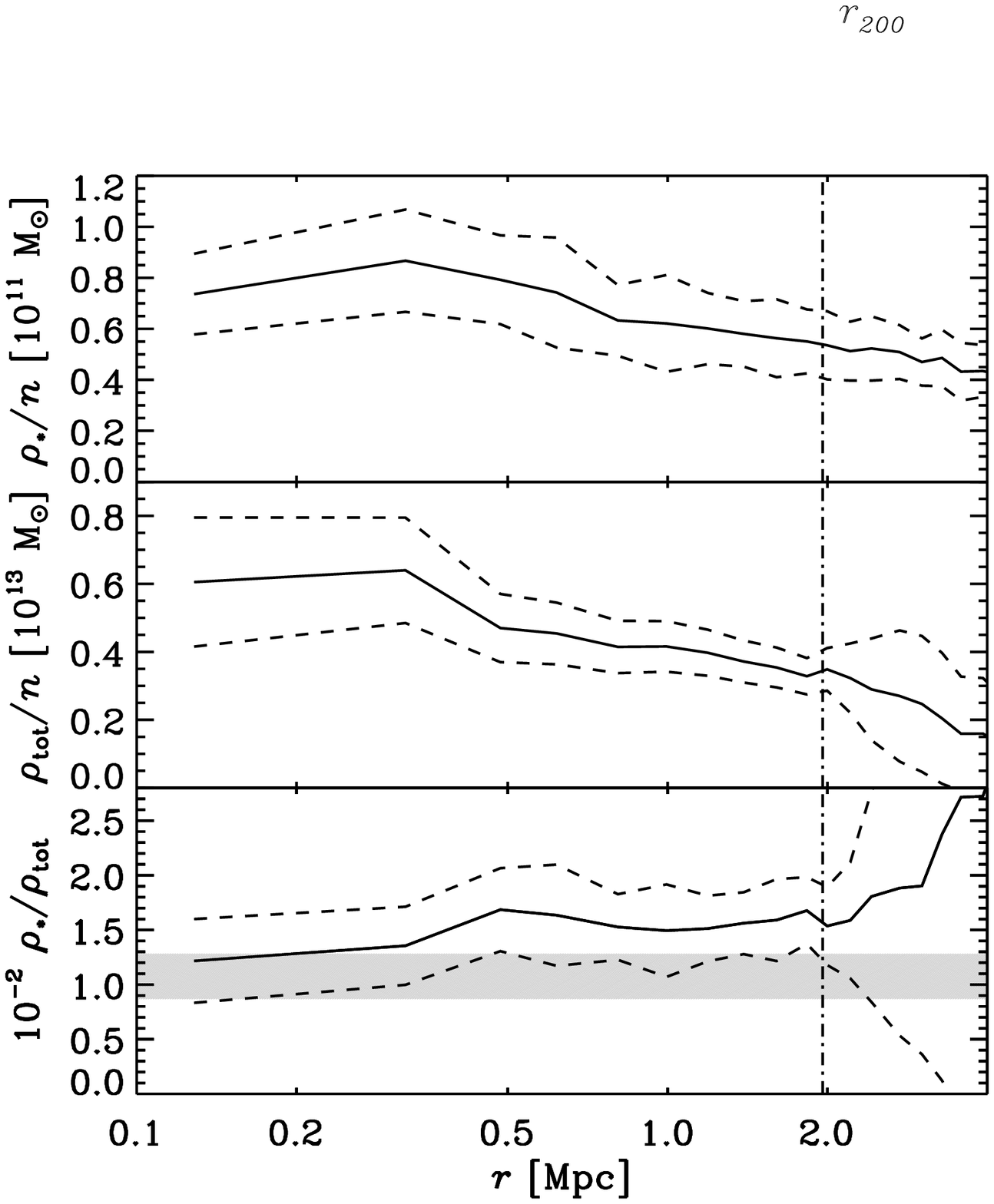}
  \caption{Top panel: the ratio of the stellar mass density and number
    density profiles. Middle panel: the ratio of the total mass
    density and number density profiles. Bottom panel: the ratio of
    the stellar mass density and total mass density profiles. Dashed
    lines indicate 1 $\sigma$ confidence regions. All densities are
    volume densities. The vertical dash-dotted line indicates the
    location of $\mathrm{r_{200}}$. The horizontal gray area
    indicates the cosmic value of the stellar mass fraction at the
    cluster mean redshift and its 1$\mathrm{\sigma}$ uncertainty.}
  \label{f:profratios}
  \end{figure}

We deproject the two density profiles using the Abel inversion, which
assumes spherical symmetry \citep[e.g.][]{BT87}. Before performing the
numerical inversion, we smooth N(R) and $\mathrm{\Sigma_{\star}(R)}$
with the LOWESS technique \citep[e.g.][]{Gebhardt+94}. The needed
extrapolation to infinity is done as in
\citet[][eq. 10]{biviano2013}. The deprojected stellar mass-to-number
density profile ratio, $\mathrm{\rho_{\star}(r)/n(r)}$, is shown in
the top panel of Fig.~\ref{f:profratios}. The dashed lines represent 1
$\sigma$ confidence levels obtained by propagation of errors, where
the fractional errors on the individual deprojected profiles are
assumed to be those estimated for the projected profiles
(Fig.~\ref{f:profs}). The ratio $\mathrm{\rho_{\star}/n}$ decreases by
$\sim 30$\% from the center to $\mathrm{r_{200}}$.

Both the relative concentration of the best-fit pNFW models of the
two projected profiles and the ratio of the two deprojected profiles,
indicate a mass segregation effect, i.e., galaxies are on average more
massive (in stars) near the cluster center than at the cluster periphery. This
is consistent with our finding that the GSNR is highest in the central
cluster region (see Fig.~\ref{f:ratio}).

We now consider the total mass density profile, $\mathrm{\rho_{tot}}$,
as given by the gravitational lensing analysis of \citet{umetsu2012}.
Specifically, we consider their NFW best-fit model parametrization of
this profile. The ratios $\mathrm{\rho_{tot}/n}$ and
$\mathrm{\rho_{\star}/\rho_{tot}}$ as a function of the 3D distance
from the cluster center, r, are shown in the middle and bottom panels
of Fig.~\ref{f:profratios}. The distribution of total mass is more
concentrated than both the distribution of galaxies \citep[see
  also][]{biviano2013} and (but less significantly so) the
distribution of stellar mass, the ratio of the stellar-to-total mass
density increasing by $\sim 20$\% from the center to
$\mathrm{r_{200}}$.  In other terms, the stellar mass fraction does
depend on radius, but this dependence is not strong.  This is
consistent with the fact that the best-fit NFW model scale radius for
the total mass density profile is only marginally different from that
of the stellar mass density profile (see Table~\ref{t:rs}).

The median value of $\mathrm{\rho_{\star}/\rho_{tot}}$ within
$\mathrm{r_{200}}$ is slightly higher (but not significantly so) than
the (physical, not comoving) cosmic value of $0.011 \pm 0.002$ at the
cluster mean redshift, evaluated using the stellar mass density values
of \citet[][their Table 2]{muzzin2013b} and our adopted cosmological
value for $\mathrm{\Omega_M}$. 

\section{Discussion}
\label{s:disc}
We find a very strong dependence of the cluster SMF on the galaxy
type.  This dependence is found in the whole cluster, as well as in
different cluster regions defined by their clustercentric distance or
by their local galaxy density. This dependence has been found
previously in several studies \citep[e.g.][]{bolzonella2010}. The sum
of the passive and SF SMFs gives rise to a SMF that deviates from a
simple Schechter beyond the $\mste$ value where the two type SMFs
cross each other \citep[see Fig.13 in][]{peng2010}. Indeed, we
find that the fit of the SMF of all galaxies by the sum of the two
best-fit Schechter functions of the passive and SF populations, is
significantly better than the fit by a single Schechter.

%%%% from the Introduction

The phenomenological model of \citeauthor{peng2010}
(\citeyear{peng2010}, well described in \citealt{baldry2012}) has
interpreted the double Schechter function shape of the galaxy SMF in
terms of `mass quenching' and `environmental quenching', which transform
star-forming (SF) galaxies into passive. If a galaxy sSFR is
independent of mass and the probability of `mass quenching' is
proportional to SFR, then the Schechter SMF of SF galaxies transforms
into a steeper single Schechter SMF of passive (quenched)
galaxies. Environmental quenching is supposed to be independent of
mass, so it does not affect the overall shape of the SMF, but only its
normalization, as SF (blue) galaxies are turned into passive (red)
galaxies. The passive SMF appears as the combination of the single
Schechter function originating from mass quenching and of another
Schechter function originating from environmental quenching.  This
bimodality should be particularly apparent in high-density regions
where the environmental quenching is most effective.  Post-quenching
mergers can then change the shape of the passive SMF by increasing the
number of very massive galaxies relative to the less massive galaxies.

This model makes specific predictions about the general evolution
  of the SMFs of SF and passive galaxies, which compare well with
  observations
  \citep{kauffmann2004,bundy2005,scoville2007,drory2008,scodeggio2009,ilbert2010,ilbert2013,huang2013,moustakas2013}. The model also predicts the differential evolution of the relative number
  density of passive and SF galaxies in different environments, which is supported by observations
  \citep{bolzonella2010,pozzetti2010}. A consequence of the
  evolutionary model of \cite{peng2010} is that the SMF for passive
galaxies should be environment dependent. Such a dependence is visible
in a local sample of galaxies (based on SDSS data,
\citealt{peng2010}), but not at higher redshift, apart from a slightly
higher density of massive galaxies in denser regions
(\citealt{bolzonella2010}). That the predicted dependence is not
  observed at high redshifts could be because of the characteristics of
  the used samples, which are generally not complete at low-masses.

%% This model predicts the observed lack of evolution of the SMF shape
%% for SF galaxies (\citealp{ilbert2010}; \citealp{huang2013}). However, a
%% decrease of the number of very massive SF galaxies with time is
%% observed (\citealt{moustakas2013}).  For field passive galaxies, the
%% model predicts an increasing low-mass end slope of the SMF with time, as
%% increasingly lower-mass SF galaxies are mass-quenched, and a nearly
%% constant massive end of the SMF with time, as mergers are rare in the
%% field, with an excess of massive galaxies in the denser environments
%% building up with time.  These predictions agree with observations (see
%% \citealt{kauffmann2004}; \citealt{scoville2007}; \citealt{scodeggio2009};
%% \citealt{ilbert2010, ilbert2013}, \citealt{moustakas2013}).

%% The model of \cite{peng2010} naturally predicts the observed increase
%% with time of the number density of red galaxies, that corresponds to a
%% decrease of the number density of blue galaxies, at a given mass
%% (\citealt{bundy2005}; \citealt{drory2008}). Since environmental
%% quenching is more effective in denser environments, this evolution is
%% predicted to be more rapid in denser environments, as observed
%% (\citealt{bolzonella2010}, \citealt{pozzetti2010}). As a consequence,
%% the SMF for passive galaxies should be environment dependent. Such a
%% dependence is visible in a local sample of galaxies (based on SDSS
%% data, \citealt{peng2010}), but not at higher redshift, apart from a
%% slightly higher density of massive galaxies in denser regions
%% (\citealt{bolzonella2010}).

%%%% end of 'from the Intro'

At the redshift of M1206 (z=0.44), the model of \citet{peng2010}
predicts that the SMFs of passive and SF galaxies should cross
  at $\mathrm{\log \mste/\msun \approx 10.1}$ in dense environments,
which is the value we find for the SMF of M1206 (see
Fig.~\ref{f:smf}). This value depends on the environment; we find the
crossing mass is $\mathrm{\log \mste/\msun \approx 10.5}$ (resp. 9.5)
for the SMF of galaxies outside (resp. within) $\mathrm{r_{200}}$, and
smaller than that found in the field by \citet[][see their
  Fig.10]{muzzin2013b}. Hence, we confirm the prediction of
\citet{peng2010} that the value above which passive galaxies dominate
the SMF shifts to lower masses in denser regions.

We find that the shape of the SMF of SF galaxies does not depend on
the environment, although we cannot examine it within the densest
cluster region for lack of statistics. This is also in line with
predictions from the model of \citet{peng2010}, and with other
observations of field galaxy SMFs
\citep[e.g.][]{ilbert2010,huang2013}.  

We also find little or no evidence of an environmental dependence of
the shape of the SMF of passive galaxies outside the very
central (densest) region.  \citet{peng2010} do predict an
environmental dependence and present evidence for it in a sample
drawn from SDSS data, but other analyses have failed to detect such a
dependence \citep[e.g.][]{vulcani2012,vulcani2013}. \citet{balogh2001}
have found the SMF of passive galaxies to be steeper in clusters than
in the field, as expected if mass quenching occurs earlier in denser
environments at a given mass \citep{peng2010}, but \citet{giodini2012}
have found that the the SMF of passive galaxies is steeper in the field than in groups of galaxies.\\
We do find a very significant change in the SMF of passive cluster
galaxies in the very inner (and densest) region, $\mathrm{R \leq 0.25
  \, r_{200}}$, corresponding to $\simeq 0.5$ Mpc (see
Fig.~\ref{f:smfs}, bottom left panel). This change corresponds to a
very steep radial decrease in the number ratio of giant
($\mathrm{\mste/\msun \geq 10^{10.5}}$) to subgiant ($\mathrm{10^{9.5}
  \leq \mste/\msun < 10^{10.5}}$) galaxies (GSNR; see
Fig.~\ref{f:ratio}), from the center to $\sim 0.8$ Mpc. Beyond this
radius the GSNR increases but more gently toward the cluster
outskirts. The GSNR of SF galaxies does not show a significant radial
dependence, but the innermost region is not sampled by our data, for
lack of a statistical significant number of SF galaxies.

Our definition of `subgiants' is close to the definition of `dwarfs'
used by \citet{sanchez-janssen2008}, i.e., galaxies 1.0 mag fainter
than the characteristic magnitude in the $r$-band luminosity function.
Their magnitude cut roughly corresponds to $\mste \sim 10^{10.5}
\msun$. Using a large number of nearby clusters they find a clear
increasing trend in the dwarf/giant number ratio with clustercentric
radius, out to $\sim 2 \, \mathrm{r_{200}}$, and they find this trend
to be due to blue galaxies, while no trend is found for the red
galaxies. Their results are therefore completely at odds with ours.
Since the cluster sample analyzed by \citet{sanchez-janssen2008}
  is at $\mathrm{z<0.1}$, this difference seems to suggest a rapid
  evolution of the GSNR, different for the different populations of
  cluster galaxies.  Quenching will transform the SF galaxies in M1206
  into passive galaxies, which could flatten the dependence of the
  passive GSNR with radius (see Fig.~\ref{f:ratio}), making it more
  similar to the GSNR observed by \citet{sanchez-janssen2008} for red
  galaxies. It is however more difficult to suggest a scenario for why
  the GSNR of blue/SF galaxies should grow a radial dependence with
  time.

Our results appear more consistent with the findings of
\citet{popesso2006} and \citet{barkhouse2009}.  \citet{popesso2006}
find a lack of dwarf, red galaxies in the central cluster
regions, and \citet{barkhouse2009} find an increase in the
number ratio of dwarf to giant red galaxies with clustercentric
radius. In both studies there is no radial trend of the blue
dwarf/giant ratio.  The comparison with our results is not
straightforward, however, as both studies are based on the analysis of
luminosity rather than mass functions. Moreover, our definition of
`subgiants' differ from their definition of `dwarfs'. In
\citet{barkhouse2009} dwarfs are galaxies that are 2.8 mag fainter
than the characteristic magnitude in the $\mathrm{R_C}$-band
luminosity function, i.e., 1.12 dex below the value of $\mathrm{M^*}$
in our SMF, or $\mste \sim 10^{9.8} \msun$. This limit is too close to
the completeness limit of our sample and we cannot adopt it as the
separation value to distinguish giant from subgiant (or dwarf)
galaxies.

An interpretation of our GSNR trends can be given in terms of a
scenario involving the processes of ram-pressure stripping
\citep{gunngott1972}, harassment \citep{moore1996,moore1998}, and
tidal destruction \citep{merritt1984}.  Ram-pressure stripping is the
process that removes a galaxy gas as it moves through the
intracluster medium.  Harassment is the cumulative effect of multiple
galaxy encounters and is able to transform a spiral galaxy into a
spheroidal galaxy. Both processes are more effective in the denser, more
central regions of a cluster. Tidal destruction is caused by the
cluster gravitational field and is effective only very close to the
cluster center \citep[e.g.][]{moran2007}.  As a SF galaxy approaches
the cluster center it is transformed into a passive galaxy by
harassment and ram pressure.  The SF galaxies are on average less massive
than passive galaxies, and in addition, their masses may become
  even smaller as they are transformed to passive galaxies, e.g., by
  harassment. As a result, the number of passive galaxies increases
  with time especially at the low-mass end, and particularly so in the
  denser cluster regions where the transformation processes are more
  effective. This creates a decreasing trend of the passive galaxy
  GSNR from the cluster outskirts to its center.  However, this trend
  might be reversed at very small radii because tidal mass
  stripping become so effective there that the low-mass galaxies are
  either totally destroyed or mass-stripped below the completeness
limit of a given survey ($\mathrm{10^{9.5} \msun}$ in our case).

A detailed and perhaps dedicated analysis of semi-analytical models in
the context of cosmological numerical simulations would be required to
(dis)prove this scenario, and this is beyond the scope of this
paper. We can however refer to the simulation work of
\citet{conselice2002} where the value $\alpha$ of the luminosity
function of cluster galaxies is first shown to increase (in absolute
values) and then decrease, as the number of interactions among
galaxies increases. The initial increase is due to tidal stripping,
until the stripping becomes so strong that stripped galaxies drop off
the completeness limit of the given survey. We do observe the same
non monotonous trend of $\alpha$ with local galaxy density (see
Sect.~\ref{sss:dens}).

Further support to our scenario comes from the comparison of
the amount of mass in the ICL and of the mass
that is missing from subgiant galaxies in the SMF of the innermost
region. To estimate the amount of mass that could have been stripped
from subgiant galaxies we proceed as follows.  First, we calculate the
total mass in the SMF of passive galaxies in the innermost Region 1
(see Sect.~\ref{sss:radial}) in the mass range between
$\mathrm{10^{9.5}-10^{10.5}\, \Msun}$, 
\begin{equation}
\mathrm{M_{sub}} \equiv \int_{10^{9.5}\, M_{\odot}}^{10^{10.5}\, M_{\odot}} \mathrm{m\, \Phi(m)\,  dm \,.}
\label{e:msub}
\end{equation}
The SMF of Region 1 shown in Fig.~\ref{f:smfs} has been
  normalized to the total number of galaxies contained in the
  subsample. In eq. \ref{e:msub} we use the non normalized SMF, with
  $\mathrm{\Phi^*=121}$. The lower limit of the integral corresponds
  to our completeness limit. The upper limit of the integral
  corresponds to the mass value where the normalized SMF of Region 1
intersects the normalized SMF of the adjacent Region 2 (see
Fig.~\ref{f:smfs}, bottom left panel).  We have chosen this mass value also to separate giant from subgiant galaxies.  We
  then recompute the integral in Eq. \ref{e:msub} by keeping the
  $\mathrm{\Phi^*}$ and $\mathrm{M^*}$ values of the SMF of Region 1
  and by changing the slope to the best-fit value found for the SMF in
  the adjacent Region 2 (see Table~\ref{t:sbf}).  The difference
between the two values of $\mathrm{M_{sub}}$ thus obtained is a
measure of how much mass is missing in the subgiant mass range in
Region 1 with respect to Region 2. We estimate the uncertainty on this
difference by repeating this estimate with slopes fixed to the $\alpha
\pm \mathrm{d} \alpha$ values of Region 2, where $\mathrm{d} \alpha$
is the error on $\alpha$ (see Table~\ref{t:sbf}). The value we find,
$\mathrm{\mathrm{\Delta M_{sub}=5.8_{-2.9}^{+3.3} \times 10^{11} \,
    \msun}}$, can be compared with the estimate of ICL stellar mass in
M1206, $\mathrm{9.9 \pm 3.8 \times 10^{11} \, \msun}$
\citep{Presotto+14}.  These two values are consistent within
$\mathrm{\sim 1 \, \sigma}$, within their admittedly large
  uncertainties. Our estimate would not change by more than 30\% if
we would extrapolate the integral of eq.~\ref{e:msub} to very low
masses, and in any case the dominant contribution to the ICL is
expected to come from intermediate-to-high mass galaxies
\citep{murante+07,contini+14}.  This comparison is consistent with a
scenario where the missing subgiant galaxies in the innermost cluster
regions have lost part of their stellar mass into a diffuse
intracluster component due to interactions with other cluster
members or with the tidal cluster field. With a very similar approach
\citet{giallongo2014} come to the same conclusions about the
nature of the ICL in another $\mathrm{z \sim 0.4}$ cluster.

The radial dependence of the GSNR is also reflected in the decreasing
stellar mass-to-number density profile ratio (see Fig.~\ref{f:ratio}).
On average, among galaxies with $\mste \geq 10^{9.5} \msun$ those near
the cluster center are $\sim 30$\% more massive than those near the
cluster virial radius. This is not because of the presence of the central
BCG, which was excluded from the analysis
when we determined the density profiles. Our finding is consistent
with the mild mass segregation found in groups by \citet{Ziparo+13},
and with the mass segregation found in clusters at $\mathrm{z \sim 1}$
by \citet{vanderburg2013}. In particular, the ratio of the best-fit
concentrations of the stellar mass density and number density profile
found by \citet{vanderburg2013}, $1.4 \pm 0.4$, is fully consistent
with the ratio we find, $1.6 \pm 0.4$. \citet{vanderburg2013} attribute
this mass segregation to dynamical friction, which should have
occurred before $\mathrm{z \sim 1}$, with little if any further
evolution thereafter.  However, mass segregation can also be the
result of tidal stripping in the central cluster region, affecting
galaxies in different ways depending on their mass. Since galaxies of lower mass
galaxies are more affected by tidal stripping, they lose mass and drop
off the completeness limit of $\mathrm{10^{9.5} \msun}$ in our
dataset.

To discriminate between dynamical friction and tidal stripping as the
driving process of mass segregation in M1206, we turn our attention to
the total mass density profile. We find that the total mass density profile is more concentrated than the galaxy number density profile, as already found by
%This is more concentrated than the
%galaxy number density profile, as already found by
\citet{biviano2013}, as well as in many other clusters
\citep[e.g.][]{BG03,LMS04,BP09}. We also find, however, that the total mass density profie is more concentrated than the stellar mass distribution, which is an entirely new result. 
%and also more concentrated than the
%stellar mass distribution, and this is an entirely new result. 
Should dynamical friction be responsible for the observed mass segregation we
would expect the total mass density profile to be less
concentrated than the stellar mass density profile, as the diffuse
dark matter component should gain energy at the expense of the
subhalos \citep[e.g.][]{DelPopolo12}. Instead, we find the
opposite. We therefore conclude that the observed mass segregation is
not due to dynamical friction in M1206, but to tidal disruption of the
less massive galaxies. Since most of the stellar mass is in the most
massive galaxies, while most of the galaxies are low-mass galaxies,
this process affects the number density profile much more severely
than the stellar mass density profile.

The radial dependence of the stellar-to-total mass ratio is very
mild. This mild dependence is consistent with the results of
\citet[][see their Table 1]{BS06}, obtained using a sample of 59
nearby clusters \citep[fully described in][]{Biviano+02}, and
\citet[][see their Fig.9]{BK14}, obtained using a sample of
$\mathrm{z<0.3}$ clusters. \citet{BK14} find that the stellar mass
fraction is roughly constant out to $\mathrm{\sim 40 \, r_{200}}$. We
confirm their result out to $\mathrm{\sim r_{200}}$; beyond that
radius our error bars become very large (see Fig.~\ref{f:ratio}).
Their determination of the average cluster stellar mass fraction also
agrees very well with the cosmic value, while our determination is
consistent, but slightly above, the cosmic value.

\section{Conclusions}
\label{s:conc}
We estimate the SMF and the stellar mass density profile,
$\mathrm{\rho_{\star}(r)}$, of the $\mathrm{z=0.44}$ cluster M1206,
using a sample of $\sim 1300$ cluster members, obtained in the
CLASH-VLT program. Cluster membership has been evaluated using
spectroscopic and photometric redshifts (for $\sim 1/3$ and $\sim 2/3$
of the members, resp.).  Stellar masses are obtained by SED fitting
with \texttt{MAGPHYS} (\citealt{dacunha2008}).  The SMF and
$\mathrm{\rho_{\star}(r)}$ are corrected for incompleteness and
contamination down to $\mathrm{\mste=10^{9.5} \msun}$. Our main
results are:
\begin{itemize}
\item The SMF of the cluster is significantly better fitted by
  a double Schechter function than by a single Schechter function. The SMFs of the
  passive and SF cluster populations are well fitted by single
  Schechter functions, with significantly different low-mass end
  slopes.
\item The SMFs of passive and SF cluster members cross at
  $\mathrm{\mste/\msun \simeq 10^{10.1}}$, in agreement with the
  prediction of the model of \citet{peng2010}. This crossing mass is
  higher in lower density regions.
\item The shape of the SMF of SF galaxies is independent from the
  environment, as defined by either the local number density of galaxies,
  or the clustercentric radius, in the range $\sim 0.8-4.0$ Mpc
  (corresponding to $\sim 0.4-2.0 \, \mathrm{r_{200}}$).
\item The shape of the SMF of passive galaxies does depend on the
  environment, since the SMF decreases more steeply toward the
  low-mass end, in the innermost cluster region ($\leq 0.5$ Mpc) than
  in the other, more external regions.
\item The number ratio of giant/subgiant galaxies is highest in the
  innermost region and lowest in the adjacent region ($0.5-1.0$
  Mpc), then the ratio increases with radius toward the cluster outskirts.
\item Both the number density and stellar mass density profiles can be
  fitted reasonably well by projected NFW models, but with different
  concentrations.  The stellar mass density profile is significantly
  more concentrated than the number density profile and only slightly
  less concentrated than the total mass density profile.
\item A possible interpretation of the environmental dependence of the
  SMF of passive galaxies and of the relative concentrations of the
  total, stellar mass, and number density profiles, is proposed in
  terms of tidal disruption of the less massive galaxies in the
  central cluster regions. On the other hand, dynamical friction seems
  not to be effective. Support for our interpretation comes from the
  comparison of the mass in the cluster ICL with the missing mass in
  the subgiant galaxy mass range in the innermost region.
\end{itemize}

In the future, we plan to extend our analysis of the SMF to the full set
of CLASH-VLT clusters (Rosati et al. in prep.) and to test our
scenario for the environmental dependence of the passive galaxy SMF
with semi-analytical models within cosmological numerical simulations.

\begin{acknowledgements}
We thank the referee for providing useful and constructive comments
that helped to improve the quality of this paper, and
Elisabete da Cunha and Adam Muzzin for useful discussions.
We also thank Elisabete da Cunha for providing us with a non public
library of models for the \texttt{MAGPHYS} SED fitting procedure.  We
acknowledge financial support from MIUR PRIN2010-2011
(J91J12000450001).  R.D. gratefully acknowledges the support provided
by the BASAL Center for Astrophysics and Associated Technologies
(CATA), and by FONDECYT grant N. 1130528. Based [in part] on data
collected at Subaru Telescope and obtained from the SMOKA, which is
operated by the Astronomy Data Center, National Astronomical
Observatory of Japan.
\end{acknowledgements}

\bibliographystyle{aa}

\bibliography{bibliography}

\begin{thebibliography}{107}
\expandafter\ifx\csname natexlab\endcsname\relax\def\natexlab#1{#1}\fi

\bibitem[{{Andreon}(2013)}]{andreon2013}
{Andreon}, S. 2013, \aap, 554, A79

\bibitem[{{Baba} {et~al.}(2002){Baba}, {Yasuda}, {Ichikawa}, {Yagi}, {Iwamoto},
  {Takata}, {Horaguchi}, {Taga}, {Watanabe}, {Ozawa}, \& {Hamabe}}]{Baba+02}
{Baba}, H., {Yasuda}, N., {Ichikawa}, S.-I., {et~al.} 2002, in Astronomical
  Society of the Pacific Conference Series, Vol. 281, Astronomical Data
  Analysis Software and Systems XI, ed. D.~A. {Bohlender}, D.~{Durand}, \&
  T.~H. {Handley}, 298

\bibitem[{{Bahcall} \& {Kulier}(2014)}]{BK14}
{Bahcall}, N.~A. \& {Kulier}, A. 2014, \mnras

\bibitem[{{Baldry} {et~al.}(2004){Baldry}, {Balogh}, {Bower}, {Glazebrook}, \&
  {Nichol}}]{baldry2004}
{Baldry}, I.~K., {Balogh}, M.~L., {Bower}, R., {Glazebrook}, K., \& {Nichol},
  R.~C. 2004, in American Institute of Physics Conference Series, Vol. 743, The
  New Cosmology: Conference on Strings and Cosmology, ed. R.~E. {Allen}, D.~V.
  {Nanopoulos}, \& C.~N. {Pope}, 106--119

\bibitem[{{Baldry} {et~al.}(2012){Baldry}, {Driver}, {Loveday}, {Taylor},
  {Kelvin}, {Liske}, {Norberg}, {Robotham}, {Brough}, {Hopkins}, {Bamford},
  {Peacock}, {Bland-Hawthorn}, {Conselice}, {Croom}, {Jones}, {Parkinson},
  {Popescu}, {Prescott}, {Sharp}, \& {Tuffs}}]{baldry2012}
{Baldry}, I.~K., {Driver}, S.~P., {Loveday}, J., {et~al.} 2012, \mnras, 421,
  621

\bibitem[{{Baldry} {et~al.}(2008){Baldry}, {Glazebrook}, \&
  {Driver}}]{baldry2008}
{Baldry}, I.~K., {Glazebrook}, K., \& {Driver}, S.~P. 2008, \mnras, 388, 945

\bibitem[{{Balogh} {et~al.}(2001){Balogh}, {Christlein}, {Zabludoff}, \&
  {Zaritsky}}]{balogh2001}
{Balogh}, M.~L., {Christlein}, D., {Zabludoff}, A.~I., \& {Zaritsky}, D. 2001,
  \apj, 557, 117

\bibitem[{{Barkhouse} {et~al.}(2009){Barkhouse}, {Yee}, \&
  {L{\'o}pez-Cruz}}]{barkhouse2009}
{Barkhouse}, W.~A., {Yee}, H.~K.~C., \& {L{\'o}pez-Cruz}, O. 2009, \apj, 703,
  2024

\bibitem[{{Bartelmann}(1996)}]{Bartelmann96}
{Bartelmann}, M. 1996, \aap, 313, 697

\bibitem[{{Bielby} {et~al.}(2012){Bielby}, {Hudelot}, {McCracken}, {Ilbert},
  {Daddi}, {Le F{\`e}vre}, {Gonzalez-Perez}, {Kneib}, {Marmo}, {Mellier},
  {Salvato}, {Sanders}, \& {Willott}}]{bielby2012}
{Bielby}, R., {Hudelot}, P., {McCracken}, H.~J., {et~al.} 2012, \aap, 545, A23

\bibitem[{{Binney} \& {Tremaine}(1987)}]{BT87}
{Binney}, J. \& {Tremaine}, S. 1987, Galactic dynamics (Princeton, NJ,
  Princeton University Press, 1987, 747 p.)

\bibitem[{{Biviano}(2008)}]{biviano2008}
{Biviano}, A. 2008, arXiv:0811.3535

\bibitem[{{Biviano} \& {Girardi}(2003)}]{BG03}
{Biviano}, A. \& {Girardi}, M. 2003, \apj, 585, 205

\bibitem[{{Biviano} {et~al.}(2002){Biviano}, {Katgert}, {Thomas}, \&
  {Adami}}]{Biviano+02}
{Biviano}, A., {Katgert}, P., {Thomas}, T., \& {Adami}, C. 2002, \aap, 387, 8

\bibitem[{{Biviano} \& {Poggianti}(2009)}]{BP09}
{Biviano}, A. \& {Poggianti}, B.~M. 2009, \aap, 501, 419

\bibitem[{{Biviano} {et~al.}(2013){Biviano}, {Rosati}, {Balestra}, {Mercurio},
  {Girardi}, {Nonino}, {Grillo}, {Scodeggio}, {Lemze}, {Kelson}, {Umetsu},
  {Postman}, {Zitrin}, {Czoske}, {Ettori}, {Fritz}, {Lombardi}, {Maier},
  {Medezinski}, {Mei}, {Presotto}, {Strazzullo}, {Tozzi}, {Ziegler},
  {Annunziatella}, {Bartelmann}, {Benitez}, {Bradley}, {Brescia}, {Broadhurst},
  {Coe}, {Demarco}, {Donahue}, {Ford}, {Gobat}, {Graves}, {Koekemoer},
  {Kuchner}, {Melchior}, {Meneghetti}, {Merten}, {Moustakas}, {Munari}, {Reg{\H
  o}s}, {Sartoris}, {Seitz}, \& {Zheng}}]{biviano2013}
{Biviano}, A., {Rosati}, P., {Balestra}, I., {et~al.} 2013, \aap, 558, A1

\bibitem[{{Biviano} \& {Salucci}(2006)}]{BS06}
{Biviano}, A. \& {Salucci}, P. 2006, \aap, 452, 75

\bibitem[{{Bolzonella} {et~al.}(2010){Bolzonella}, {Kova{\v c}}, {Pozzetti},
  {Zucca}, {Cucciati}, {Lilly}, {Peng}, {Iovino}, {Zamorani}, {Vergani},
  {Tasca}, {Lamareille}, {Oesch}, {Caputi}, {Kampczyk}, {Bardelli}, {Maier},
  {Abbas}, {Knobel}, {Scodeggio}, {Carollo}, {Contini}, {Kneib}, {Le
  F{\`e}vre}, {Mainieri}, {Renzini}, {Bongiorno}, {Coppa}, {de la Torre}, {de
  Ravel}, {Franzetti}, {Garilli}, {Le Borgne}, {Le Brun}, {Mignoli},
  {Pell{\'o}}, {Perez-Montero}, {Ricciardelli}, {Silverman}, {Tanaka},
  {Tresse}, {Bottini}, {Cappi}, {Cassata}, {Cimatti}, {Guzzo}, {Koekemoer},
  {Leauthaud}, {Maccagni}, {Marinoni}, {McCracken}, {Memeo}, {Meneux},
  {Porciani}, {Scaramella}, {Aussel}, {Capak}, {Halliday}, {Ilbert},
  {Kartaltepe}, {Salvato}, {Sanders}, {Scarlata}, {Scoville}, {Taniguchi}, \&
  {Thompson}}]{bolzonella2010}
{Bolzonella}, M., {Kova{\v c}}, K., {Pozzetti}, L., {et~al.} 2010, \aap, 524,
  A76

\bibitem[{{Brammer} {et~al.}(2008){Brammer}, {van Dokkum}, \&
  {Coppi}}]{brammer2008}
{Brammer}, G.~B., {van Dokkum}, P.~G., \& {Coppi}, P. 2008, \apj, 686, 1503

\bibitem[{{Brescia} {et~al.}(2013){Brescia}, {Cavuoti}, {D'Abrusco}, {Longo},
  \& {Mercurio}}]{brescia2013}
{Brescia}, M., {Cavuoti}, S., {D'Abrusco}, R., {Longo}, G., \& {Mercurio}, A.
  2013, \apj, 772, 140

\bibitem[{{Bruzual} \& {Charlot}(2003)}]{bruzual&charlot2003}
{Bruzual}, G. \& {Charlot}, S. 2003, \mnras, 344, 1000

\bibitem[{{Bundy} {et~al.}(2005){Bundy}, {Ellis}, \& {Conselice}}]{bundy2005}
{Bundy}, K., {Ellis}, R.~S., \& {Conselice}, C.~J. 2005, \apj, 625, 621

\bibitem[{{Calvi} {et~al.}(2013){Calvi}, {Poggianti}, {Vulcani}, \&
  {Fasano}}]{calvi2013}
{Calvi}, R., {Poggianti}, B.~M., {Vulcani}, B., \& {Fasano}, G. 2013, \mnras,
  432, 3141

\bibitem[{{Capozzi} {et~al.}(2012){Capozzi}, {Collins}, {Stott}, \&
  {Hilton}}]{capozzi2012}
{Capozzi}, D., {Collins}, C.~A., {Stott}, J.~P., \& {Hilton}, M. 2012, \mnras,
  419, 2821

\bibitem[{{Chabrier}(2003)}]{chabrier2003}
{Chabrier}, G. 2003, \pasp, 115, 763

\bibitem[{{Charlot} \& {Fall}(2000)}]{CF00}
{Charlot}, S. \& {Fall}, S.~M. 2000, \apj, 539, 718

\bibitem[{{Conselice}(2002)}]{conselice2002}
{Conselice}, C.~J. 2002, \apjl, 573, L5

\bibitem[{{Contini} {et~al.}(2014){Contini}, {De Lucia}, {Villalobos}, \&
  {Borgani}}]{contini+14}
{Contini}, E., {De Lucia}, G., {Villalobos}, {\'A}., \& {Borgani}, S. 2014,
  \mnras, 437, 3787

\bibitem[{{Cucciati} {et~al.}(2012){Cucciati}, {De Lucia}, {Zucca}, {Iovino},
  {de la Torre}, {Pozzetti}, {Blaizot}, {Zamorani}, {Bolzonella}, {Vergani},
  {Bardelli}, {Tresse}, \& {Pollo}}]{cucciati2012}
{Cucciati}, O., {De Lucia}, G., {Zucca}, E., {et~al.} 2012, \aap, 548, A108

\bibitem[{{da Cunha} {et~al.}(2008){da Cunha}, {Charlot}, \&
  {Elbaz}}]{dacunha2008}
{da Cunha}, E., {Charlot}, S., \& {Elbaz}, D. 2008, \mnras, 388, 1595

\bibitem[{{Davidzon} {et~al.}(2013){Davidzon}, {Bolzonella}, {Coupon},
  {Ilbert}, {Arnouts}, {de la Torre}, {Fritz}, {De Lucia}, {Iovino}, {Granett},
  {Zamorani}, {Guzzo}, {Abbas}, {Adami}, {Bel}, {Bottini}, {Branchini},
  {Cappi}, {Cucciati}, {Franzetti}, {Fumana}, {Garilli}, {Krywult}, {Le Brun},
  {Le F{\`e}vre}, {Maccagni}, {Ma{\l}ek}, {Marulli}, {McCracken}, {Paioro},
  {Peacock}, {Polletta}, {Pollo}, {Schlagenhaufer}, {Scodeggio}, {Tasca},
  {Tojeiro}, {Vergani}, {Zanichelli}, {Burden}, {Di Porto}, {Marchetti},
  {Marinoni}, {Mellier}, {Moscardini}, {Moutard}, {Nichol}, {Percival},
  {Phleps}, \& {Wolk}}]{Davidzon+13}
{Davidzon}, I., {Bolzonella}, M., {Coupon}, J., {et~al.} 2013, \aap, 558, A23

\bibitem[{{De Lucia} {et~al.}(2012){De Lucia}, {Weinmann}, {Poggianti},
  {Arag{\'o}n-Salamanca}, \& {Zaritsky}}]{delucia2012}
{De Lucia}, G., {Weinmann}, S., {Poggianti}, B.~M., {Arag{\'o}n-Salamanca}, A.,
  \& {Zaritsky}, D. 2012, \mnras, 423, 1277

\bibitem[{{De Propris} \& {Christlein}(2009)}]{depropris2009}
{De Propris}, R. \& {Christlein}, D. 2009, Astronomische Nachrichten, 330, 943

\bibitem[{{De Propris} {et~al.}(2007){De Propris}, {Stanford}, {Eisenhardt},
  {Holden}, \& {Rosati}}]{depropris2007}
{De Propris}, R., {Stanford}, S.~A., {Eisenhardt}, P.~R., {Holden}, B.~P., \&
  {Rosati}, P. 2007, \aj, 133, 2209

\bibitem[{{Del Popolo}(2012)}]{DelPopolo12}
{Del Popolo}, A. 2012, \mnras, 424, 38

\bibitem[{{Dressler}(1980)}]{dressler1980}
{Dressler}, A. 1980, \apj, 236, 351

\bibitem[{{Drory} \& {Alvarez}(2008)}]{drory2008}
{Drory}, N. \& {Alvarez}, M. 2008, \apj, 680, 41

\bibitem[{{Ebeling} {et~al.}(2001){Ebeling}, {Edge}, \& {Henry}}]{Ebeling+01}
{Ebeling}, H., {Edge}, A.~C., \& {Henry}, J.~P. 2001, \apj, 553, 668

\bibitem[{{Ebeling} {et~al.}(2009{\natexlab{a}}){Ebeling}, {Ma}, {Kneib},
  {Jullo}, {Courtney}, {Barrett}, {Edge}, \& {Le Borgne}}]{ebeling2009}
{Ebeling}, H., {Ma}, C.~J., {Kneib}, J.-P., {et~al.} 2009{\natexlab{a}},
  \mnras, 395, 1213

\bibitem[{{Ebeling} {et~al.}(2009{\natexlab{b}}){Ebeling}, {Ma}, {Kneib},
  {Jullo}, {Courtney}, {Barrett}, {Edge}, \& {Le Borgne}}]{Ebeling+09}
{Ebeling}, H., {Ma}, C.~J., {Kneib}, J.-P., {et~al.} 2009{\natexlab{b}},
  \mnras, 395, 1213

\bibitem[{{Fontana} {et~al.}(2006){Fontana}, {Salimbeni}, {Grazian},
  {Giallongo}, {Pentericci}, {Nonino}, {Fontanot}, {Menci}, {Monaco},
  {Cristiani}, {Vanzella}, {de Santis}, \& {Gallozzi}}]{fontana2006}
{Fontana}, A., {Salimbeni}, S., {Grazian}, A., {et~al.} 2006, \aap, 459, 745

\bibitem[{{Gebhardt} {et~al.}(1994){Gebhardt}, {Pryor}, {Williams}, \&
  {Hesser}}]{Gebhardt+94}
{Gebhardt}, K., {Pryor}, C., {Williams}, T.~B., \& {Hesser}, J.~E. 1994, \aj,
  107, 2067

\bibitem[{{Giallongo} {et~al.}(2014){Giallongo}, {Menci}, {Grazian},
  {Gallozzi}, {Castellano}, {Fiore}, {Fontana}, {Pentericci}, {Boutsia},
  {Paris}, {Speziali}, \& {Testa}}]{giallongo2014}
{Giallongo}, E., {Menci}, N., {Grazian}, A., {et~al.} 2014, \apj, 781, 24

\bibitem[{{Giodini} {et~al.}(2012){Giodini}, {Finoguenov}, {Pierini},
  {Zamorani}, {Ilbert}, {Lilly}, {Peng}, {Scoville}, \& {Tanaka}}]{giodini2012}
{Giodini}, S., {Finoguenov}, A., {Pierini}, D., {et~al.} 2012, \aap, 538, A104

\bibitem[{{Gunn} \& {Gott}(1972)}]{gunngott1972}
{Gunn}, J.~E. \& {Gott}, III, J.~R. 1972, \apj, 176, 1

\bibitem[{{Huang} {et~al.}(2013){Huang}, {Faber}, {Willmer}, {Rigopoulou},
  {Koo}, {Newman}, {Shu}, {Ashby}, {Barmby}, {Coil}, {Luo}, {Magdis}, {Wang},
  {Weiner}, {Willner}, {Zheng}, \& {Fazio}}]{huang2013}
{Huang}, J.-S., {Faber}, S.~M., {Willmer}, C.~N.~A., {et~al.} 2013, \apj, 766,
  21

\bibitem[{{Ilbert} {et~al.}(2013){Ilbert}, {McCracken}, {Le F{\`e}vre},
  {Capak}, {Dunlop}, {Karim}, {Renzini}, {Caputi}, {Boissier}, {Arnouts},
  {Aussel}, {Comparat}, {Guo}, {Hudelot}, {Kartaltepe}, {Kneib}, {Krogager},
  {Le Floc'h}, {Lilly}, {Mellier}, {Milvang-Jensen}, {Moutard}, {Onodera},
  {Richard}, {Salvato}, {Sanders}, {Scoville}, {Silverman}, {Taniguchi},
  {Tasca}, {Thomas}, {Toft}, {Tresse}, {Vergani}, {Wolk}, \&
  {Zirm}}]{ilbert2013}
{Ilbert}, O., {McCracken}, H.~J., {Le F{\`e}vre}, O., {et~al.} 2013, \aap, 556,
  A55

\bibitem[{{Ilbert} {et~al.}(2010){Ilbert}, {Salvato}, {Le Floc'h}, {Aussel},
  {Capak}, {McCracken}, {Mobasher}, {Kartaltepe}, {Scoville}, {Sanders},
  {Arnouts}, {Bundy}, {Cassata}, {Kneib}, {Koekemoer}, {Le F{\`e}vre}, {Lilly},
  {Surace}, {Taniguchi}, {Tasca}, {Thompson}, {Tresse}, {Zamojski}, {Zamorani},
  \& {Zucca}}]{ilbert2010}
{Ilbert}, O., {Salvato}, M., {Le Floc'h}, E., {et~al.} 2010, \apj, 709, 644

\bibitem[{{Jones} {et~al.}(2004){Jones}, {Saunders}, {Colless}, {Read},
  {Parker}, {Watson}, {Campbell}, {Burkey}, {Mauch}, {Moore}, {Hartley},
  {Cass}, {James}, {Russell}, {Fiegert}, {Dawe}, {Huchra}, {Jarrett}, {Lahav},
  {Lucey}, {Mamon}, {Proust}, {Sadler}, \& {Wakamatsu}}]{Jones+04}
{Jones}, D.~H., {Saunders}, W., {Colless}, M., {et~al.} 2004, \mnras, 355, 747

\bibitem[{{Kauffmann} {et~al.}(2004){Kauffmann}, {White}, {Heckman},
  {M{\'e}nard}, {Brinchmann}, {Charlot}, {Tremonti}, \&
  {Brinkmann}}]{kauffmann2004}
{Kauffmann}, G., {White}, S.~D.~M., {Heckman}, T.~M., {et~al.} 2004, \mnras,
  353, 713

\bibitem[{Kauffmann {et~al.}(2003)}]{kauffmann2003}
Kauffmann, G. {et~al.} 2003, Mon.Not.Roy.Astron.Soc., 341, 33

\bibitem[{{Kodama} \& {Bower}(2003)}]{kodama&bower2003}
{Kodama}, T. \& {Bower}, R. 2003, \mnras, 346, 1

\bibitem[{{Kriek} {et~al.}(2010){Kriek}, {Labb{\'e}}, {Conroy}, {Whitaker},
  {van Dokkum}, {Brammer}, {Franx}, {Illingworth}, {Marchesini}, {Muzzin},
  {Quadri}, \& {Rudnick}}]{kriek2010}
{Kriek}, M., {Labb{\'e}}, I., {Conroy}, C., {et~al.} 2010, \apjl, 722, L64

\bibitem[{{Kriek} {et~al.}(2009){Kriek}, {van Dokkum}, {Labb{\'e}}, {Franx},
  {Illingworth}, {Marchesini}, \& {Quadri}}]{kriek2009}
{Kriek}, M., {van Dokkum}, P.~G., {Labb{\'e}}, I., {et~al.} 2009, \apj, 700,
  221

\bibitem[{{Lamareille} {et~al.}(2006){Lamareille}, {Contini}, {Le Borgne},
  {Brinchmann}, {Charlot}, \& {Richard}}]{Lamareille+06}
{Lamareille}, F., {Contini}, T., {Le Borgne}, J.-F., {et~al.} 2006, \aap, 448,
  893

\bibitem[{{Lara-L{\'o}pez} {et~al.}(2010){Lara-L{\'o}pez}, {Bongiovanni},
  {Cepa}, {P{\'e}rez Garc{\'{\i}}a}, {S{\'a}nchez-Portal}, {Casta{\~n}eda},
  {Fern{\'a}ndez Lorenzo}, \& {Povi{\'c}}}]{laralopez2010}
{Lara-L{\'o}pez}, M.~A., {Bongiovanni}, A., {Cepa}, J., {et~al.} 2010, \aap,
  519, A31

\bibitem[{{Le F{\`e}vre} {et~al.}(2003){Le F{\`e}vre}, {Saisse}, {Mancini},
  {Brau-Nogue}, {Caputi}, {Castinel}, {D'Odorico}, {Garilli}, {Kissler-Patig},
  {Lucuix}, {Mancini}, {Pauget}, {Sciarretta}, {Scodeggio}, {Tresse}, \&
  {Vettolani}}]{LeFevre+03}
{Le F{\`e}vre}, O., {Saisse}, M., {Mancini}, D., {et~al.} 2003, in Society of
  Photo-Optical Instrumentation Engineers (SPIE) Conference Series, Vol. 4841,
  Society of Photo-Optical Instrumentation Engineers (SPIE) Conference Series,
  ed. M.~{Iye} \& A.~F.~M. {Moorwood}, 1670--1681

\bibitem[{{Lemze} {et~al.}(2013){Lemze}, {Postman}, {Genel}, {Ford},
  {Balestra}, {Donahue}, {Kelson}, {Nonino}, {Mercurio}, {Biviano}, {Rosati},
  {Umetsu}, {Sand}, {Koekemoer}, {Meneghetti}, {Melchior}, {Newman}, {Bhatti},
  {Voit}, {Medezinski}, {Zitrin}, {Zheng}, {Broadhurst}, {Bartelmann},
  {Benitez}, {Bouwens}, {Bradley}, {Coe}, {Graves}, {Grillo}, {Infante},
  {Jimenez-Teja}, {Jouvel}, {Lahav}, {Maoz}, {Merten}, {Molino}, {Moustakas},
  {Moustakas}, {Ogaz}, {Scodeggio}, \& {Seitz}}]{Lemze2013}
{Lemze}, D., {Postman}, M., {Genel}, S., {et~al.} 2013, \apj, 776, 91

\bibitem[{{Lin} {et~al.}(2006){Lin}, {Mohr}, {Gonzalez}, \&
  {Stanford}}]{lin2006}
{Lin}, Y.-T., {Mohr}, J.~J., {Gonzalez}, A.~H., \& {Stanford}, S.~A. 2006,
  \apjl, 650, L99

\bibitem[{{Lin} {et~al.}(2004{\natexlab{a}}){Lin}, {Mohr}, \&
  {Stanford}}]{lin2004}
{Lin}, Y.-T., {Mohr}, J.~J., \& {Stanford}, S.~A. 2004{\natexlab{a}}, \apj,
  610, 745

\bibitem[{{Lin} {et~al.}(2004{\natexlab{b}}){Lin}, {Mohr}, \&
  {Stanford}}]{LMS04}
{Lin}, Y.-T., {Mohr}, J.~J., \& {Stanford}, S.~A. 2004{\natexlab{b}}, \apj,
  610, 745

\bibitem[{{Macci{\`o}} {et~al.}(2008){Macci{\`o}}, {Dutton}, \& {van den
  Bosch}}]{MDvdB08}
{Macci{\`o}}, A.~V., {Dutton}, A.~A., \& {van den Bosch}, F.~C. 2008, \mnras,
  391, 1940

\bibitem[{{Macci{\`o}} {et~al.}(2010){Macci{\`o}}, {Kang}, {Fontanot},
  {Somerville}, {Koposov}, \& {Monaco}}]{maccio2010}
{Macci{\`o}}, A.~V., {Kang}, X., {Fontanot}, F., {et~al.} 2010, \mnras, 402,
  1995

\bibitem[{{Malumuth} \& {Kriss}(1986)}]{MK86}
{Malumuth}, E.~M. \& {Kriss}, G.~A. 1986, \apj, 308, 10

\bibitem[{{Mamon} {et~al.}(2013){Mamon}, {Biviano}, \& {Bou{\'e}}}]{mamon2013}
{Mamon}, G.~A., {Biviano}, A., \& {Bou{\'e}}, G. 2013, \mnras, 429, 3079

\bibitem[{{Mamon} {et~al.}(2010){Mamon}, {Biviano}, \& {Murante}}]{MBM10}
{Mamon}, G.~A., {Biviano}, A., \& {Murante}, G. 2010, \aap, 520, A30

\bibitem[{{Mancone} {et~al.}(2012){Mancone}, {Baker}, {Gonzalez}, {Ashby},
  {Stanford}, {Brodwin}, {Eisenhardt}, {Snyder}, {Stern}, \&
  {Wright}}]{mancone2012}
{Mancone}, C.~L., {Baker}, T., {Gonzalez}, A.~H., {et~al.} 2012, \apj, 761, 141

\bibitem[{{Maraston} {et~al.}(2006){Maraston}, {Daddi}, {Renzini}, {Cimatti},
  {Dickinson}, {Papovich}, {Pasquali}, \& {Pirzkal}}]{maraston2006}
{Maraston}, C., {Daddi}, E., {Renzini}, A., {et~al.} 2006, \apj, 652, 85

\bibitem[{{McCracken} {et~al.}(2013){McCracken}, {Milvang-Jensen}, {Dunlop},
  {Franx}, {Fynbo}, {Le F{\`e}vre}, {Holt}, {Caputi}, {Goranova}, {Buitrago},
  {Emerson}, {Freudling}, {Herent}, {Hudelot}, {L{\'o}pez-Sanjuan}, {Magnard},
  {Muzzin}, {Mellier}, {M{\o}ller}, {Nilsson}, {Sutherland}, {Tasca}, \&
  {Zabl}}]{mckracken2013}
{McCracken}, H.~J., {Milvang-Jensen}, B., {Dunlop}, J., {et~al.} 2013, The
  Messenger, 154, 29

\bibitem[{{Menci} {et~al.}(2012){Menci}, {Fiore}, \& {Lamastra}}]{menci2012}
{Menci}, N., {Fiore}, F., \& {Lamastra}, A. 2012, \mnras, 421, 2384

\bibitem[{{Merluzzi} {et~al.}(2010){Merluzzi}, {Mercurio}, {Haines}, {Smith},
  {Busarello}, \& {Lucey}}]{merluzzi2010}
{Merluzzi}, P., {Mercurio}, A., {Haines}, C.~P., {et~al.} 2010, \mnras, 402,
  753

\bibitem[{{Merritt}(1984)}]{merritt1984}
{Merritt}, D. 1984, \apj, 276, 26

\bibitem[{{Meyer}(1975)}]{Meyer+75}
{Meyer}, S.~L. 1975, Data Analysis for Scientists and Engineers (New York: John
  Wiley \& Sons Inc.)

\bibitem[{{Moore} {et~al.}(1998){Moore}, {Governato}, {Quinn}, {Stadel}, \&
  {Lake}}]{moore1998}
{Moore}, B., {Governato}, F., {Quinn}, T., {Stadel}, J., \& {Lake}, G. 1998,
  \apjl, 499, L5

\bibitem[{{Moore} {et~al.}(1996){Moore}, {Katz}, {Lake}, {Dressler}, \&
  {Oemler}}]{moore1996}
{Moore}, B., {Katz}, N., {Lake}, G., {Dressler}, A., \& {Oemler}, A. 1996,
  \nat, 379, 613

\bibitem[{{Moran} {et~al.}(2007){Moran}, {Ellis}, {Treu}, {Smith}, {Rich}, \&
  {Smail}}]{moran2007}
{Moran}, S.~M., {Ellis}, R.~S., {Treu}, T., {et~al.} 2007, \apj, 671, 1503

\bibitem[{{Mortlock} {et~al.}(2011){Mortlock}, {Conselice}, {Bluck}, {Bauer},
  {Gr{\"u}tzbauch}, {Buitrago}, \& {Ownsworth}}]{mortlock2011}
{Mortlock}, A., {Conselice}, C.~J., {Bluck}, A.~F.~L., {et~al.} 2011, \mnras,
  413, 2845

\bibitem[{{Moustakas} {et~al.}(2013){Moustakas}, {Coil}, {Aird}, {Blanton},
  {Cool}, {Eisenstein}, {Mendez}, {Wong}, {Zhu}, \& {Arnouts}}]{moustakas2013}
{Moustakas}, J., {Coil}, A.~L., {Aird}, J., {et~al.} 2013, \apj, 767, 50

\bibitem[{{Murante} {et~al.}(2007){Murante}, {Giovalli}, {Gerhard},
  {Arnaboldi}, {Borgani}, \& {Dolag}}]{murante+07}
{Murante}, G., {Giovalli}, M., {Gerhard}, O., {et~al.} 2007, \mnras, 377, 2

\bibitem[{{Muzzin} {et~al.}(2013{\natexlab{a}}){Muzzin}, {Marchesini},
  {Stefanon}, {Franx}, {McCracken}, {Milvang-Jensen}, {Dunlop}, {Fynbo},
  {Brammer}, {Labb{\'e}}, \& {van Dokkum}}]{muzzin2013b}
{Muzzin}, A., {Marchesini}, D., {Stefanon}, M., {et~al.} 2013{\natexlab{a}},
  \apj, 777, 18

\bibitem[{{Muzzin} {et~al.}(2013{\natexlab{b}}){Muzzin}, {Marchesini},
  {Stefanon}, {Franx}, {Milvang-Jensen}, {Dunlop}, {Fynbo}, {Brammer},
  {Labb{\'e}}, \& {van Dokkum}}]{muzzin2013a}
{Muzzin}, A., {Marchesini}, D., {Stefanon}, M., {et~al.} 2013{\natexlab{b}},
  \apjs, 206, 8

\bibitem[{{Muzzin} {et~al.}(2008){Muzzin}, {Wilson}, {Lacy}, {Yee}, \&
  {Stanford}}]{muzzin2008}
{Muzzin}, A., {Wilson}, G., {Lacy}, M., {Yee}, H.~K.~C., \& {Stanford}, S.~A.
  2008, \apj, 686, 966

\bibitem[{{Muzzin} {et~al.}(2007){Muzzin}, {Yee}, {Hall}, {Ellingson}, \&
  {Lin}}]{muzzin2007}
{Muzzin}, A., {Yee}, H.~K.~C., {Hall}, P.~B., {Ellingson}, E., \& {Lin}, H.
  2007, \apj, 659, 1106

\bibitem[{{Navarro} {et~al.}(1997){Navarro}, {Frenk}, \& {White}}]{NFW97}
{Navarro}, J.~F., {Frenk}, C.~S., \& {White}, S. D.~M. 1997, \apj, 490, 493

\bibitem[{{Peng} {et~al.}(2010){Peng}, {Lilly}, {Kova{\v c}}, {Bolzonella},
  {Pozzetti}, {Renzini}, {Zamorani}, {Ilbert}, {Knobel}, {Iovino}, {Maier},
  {Cucciati}, {Tasca}, {Carollo}, {Silverman}, {Kampczyk}, {de Ravel},
  {Sanders}, {Scoville}, {Contini}, {Mainieri}, {Scodeggio}, {Kneib}, {Le
  F{\`e}vre}, {Bardelli}, {Bongiorno}, {Caputi}, {Coppa}, {de la Torre},
  {Franzetti}, {Garilli}, {Lamareille}, {Le Borgne}, {Le Brun}, {Mignoli},
  {Perez Montero}, {Pello}, {Ricciardelli}, {Tanaka}, {Tresse}, {Vergani},
  {Welikala}, {Zucca}, {Oesch}, {Abbas}, {Barnes}, {Bordoloi}, {Bottini},
  {Cappi}, {Cassata}, {Cimatti}, {Fumana}, {Hasinger}, {Koekemoer},
  {Leauthaud}, {Maccagni}, {Marinoni}, {McCracken}, {Memeo}, {Meneux}, {Nair},
  {Porciani}, {Presotto}, \& {Scaramella}}]{peng2010}
{Peng}, Y.-j., {Lilly}, S.~J., {Kova{\v c}}, K., {et~al.} 2010, \apj, 721, 193

\bibitem[{{Popesso} {et~al.}(2006){Popesso}, {Biviano}, {B{\"o}hringer}, \&
  {Romaniello}}]{popesso2006}
{Popesso}, P., {Biviano}, A., {B{\"o}hringer}, H., \& {Romaniello}, M. 2006,
  \aap, 445, 29

\bibitem[{{Postman} {et~al.}(2012){Postman}, {Coe}, {Ben{\'{\i}}tez},
  {Bradley}, {Broadhurst}, {Donahue}, {Ford}, {Graur}, {Graves}, {Jouvel},
  {Koekemoer}, {Lemze}, {Medezinski}, {Molino}, {Moustakas}, {Ogaz}, {Riess},
  {Rodney}, {Rosati}, {Umetsu}, {Zheng}, {Zitrin}, {Bartelmann}, {Bouwens},
  {Czakon}, {Golwala}, {Host}, {Infante}, {Jha}, {Jimenez-Teja}, {Kelson},
  {Lahav}, {Lazkoz}, {Maoz}, {McCully}, {Melchior}, {Meneghetti}, {Merten},
  {Moustakas}, {Nonino}, {Patel}, {Reg{\"o}s}, {Sayers}, {Seitz}, \& {Van der
  Wel}}]{postman2012}
{Postman}, M., {Coe}, D., {Ben{\'{\i}}tez}, N., {et~al.} 2012, \apjs, 199, 25

\bibitem[{{Pozzetti} {et~al.}(2010){Pozzetti}, {Bolzonella}, {Zucca},
  {Zamorani}, {Lilly}, {Renzini}, {Moresco}, {Mignoli}, {Cassata}, {Tasca},
  {Lamareille}, {Maier}, {Meneux}, {Halliday}, {Oesch}, {Vergani}, {Caputi},
  {Kova{\v c}}, {Cimatti}, {Cucciati}, {Iovino}, {Peng}, {Carollo}, {Contini},
  {Kneib}, {Le F{\'e}vre}, {Mainieri}, {Scodeggio}, {Bardelli}, {Bongiorno},
  {Coppa}, {de la Torre}, {de Ravel}, {Franzetti}, {Garilli}, {Kampczyk},
  {Knobel}, {Le Borgne}, {Le Brun}, {Pell{\`o}}, {Perez Montero},
  {Ricciardelli}, {Silverman}, {Tanaka}, {Tresse}, {Abbas}, {Bottini}, {Cappi},
  {Guzzo}, {Koekemoer}, {Leauthaud}, {Maccagni}, {Marinoni}, {McCracken},
  {Memeo}, {Porciani}, {Scaramella}, {Scarlata}, \& {Scoville}}]{pozzetti2010}
{Pozzetti}, L., {Bolzonella}, M., {Zucca}, E., {et~al.} 2010, \aap, 523, A13

\bibitem[{{Presotto} {et~al.}(2014){Presotto}, {Girardi}, {Nonino}, {Mercurio},
  {Grillo}, {Rosati}, {Biviano}, {Annunziatella}, {Balestra}, {Cui},
  {Sartoris}, {Lemze}, {Ascaso}, {Moustakas}, {Ford}, {Fritz}, {Czoske},
  {Ettori}, {Kuchner}, {Lombardi}, {Maier}, {Medezinski}, {Molino},
  {Scodeggio}, {Strazzullo}, {Tozzi}, {Ziegler}, {Bartelmann}, {Benitez},
  {Bradley}, {Brescia}, {Broadhurst}, {Coe}, {Donahue}, {Gobat}, {Graves},
  {Kelson}, {Koekemoer}, {Melchior}, {Meneghetti}, {Merten}, {Moustakas},
  {Munari}, {Postman}, {Reg{\H o}s}, {Seitz}, {Umetsu}, {Zheng}, \&
  {Zitrin}}]{Presotto+14}
{Presotto}, V., {Girardi}, M., {Nonino}, M., {et~al.} 2014, arXiv:1403.4979

\bibitem[{{Press} {et~al.}(1993){Press}, {Rybicki}, \& {Schneider}}]{press2013}
{Press}, W.~H., {Rybicki}, G.~B., \& {Schneider}, D.~P. 1993, \apj, 414, 64

\bibitem[{{S{\'a}nchez-Janssen} {et~al.}(2008){S{\'a}nchez-Janssen}, {Aguerri},
  \& {Mu{\~n}oz-Tu{\~n}{\'o}n}}]{sanchez-janssen2008}
{S{\'a}nchez-Janssen}, R., {Aguerri}, J.~A.~L., \& {Mu{\~n}oz-Tu{\~n}{\'o}n},
  C. 2008, \apjl, 679, L77

\bibitem[{{Schechter}(1976)}]{schechter1976}
{Schechter}, P. 1976, \apj, 203, 297

\bibitem[{{Scodeggio} {et~al.}(2005){Scodeggio}, {Franzetti}, {Garilli},
  {Zanichelli}, {Paltani}, {Maccagni}, {Bottini}, {Le Brun}, {Contini},
  {Scaramella}, {Adami}, {Bardelli}, {Zucca}, {Tresse}, {Ilbert}, {Foucaud},
  {Iovino}, {Merighi}, {Zamorani}, {Gavignaud}, {Rizzo}, {McCracken}, {Le
  F{\`e}vre}, {Picat}, {Vettolani}, {Arnaboldi}, {Arnouts}, {Bolzonella},
  {Cappi}, {Charlot}, {Ciliegi}, {Guzzo}, {Marano}, {Marinoni}, {Mathez},
  {Mazure}, {Meneux}, {Pell{\`o}}, {Pollo}, {Pozzetti}, \&
  {Radovich}}]{Scodeggio+05}
{Scodeggio}, M., {Franzetti}, P., {Garilli}, B., {et~al.} 2005, \pasp, 117,
  1284

\bibitem[{{Scodeggio} {et~al.}(2009){Scodeggio}, {Vergani}, {Cucciati},
  {Iovino}, {Franzetti}, {Garilli}, {Lamareille}, {Bolzonella}, {Pozzetti},
  {Abbas}, {Marinoni}, {Contini}, {Bottini}, {Le Brun}, {Le F{\`e}vre},
  {Maccagni}, {Scaramella}, {Tresse}, {Vettolani}, {Zanichelli}, {Adami},
  {Arnouts}, {Bardelli}, {Cappi}, {Charlot}, {Ciliegi}, {Foucaud}, {Gavignaud},
  {Guzzo}, {Ilbert}, {McCracken}, {Marano}, {Mazure}, {Meneux}, {Merighi},
  {Paltani}, {Pell{\`o}}, {Pollo}, {Radovich}, {Zamorani}, {Zucca}, {Bondi},
  {Bongiorno}, {Brinchmann}, {de La Torre}, {de Ravel}, {Gregorini}, {Memeo},
  {Perez-Montero}, {Mellier}, {Temporin}, \& {Walcher}}]{scodeggio2009}
{Scodeggio}, M., {Vergani}, D., {Cucciati}, O., {et~al.} 2009, \aap, 501, 21

\bibitem[{{Scoville} {et~al.}(2007){Scoville}, {Aussel}, {Benson}, {Blain},
  {Calzetti}, {Capak}, {Ellis}, {El-Zant}, {Finoguenov}, {Giavalisco}, {Guzzo},
  {Hasinger}, {Koda}, {Le F{\`e}vre}, {Massey}, {McCracken}, {Mobasher},
  {Renzini}, {Rhodes}, {Salvato}, {Sanders}, {Sasaki}, {Schinnerer}, {Sheth},
  {Shopbell}, {Taniguchi}, {Taylor}, \& {Thompson}}]{scoville2007}
{Scoville}, N., {Aussel}, H., {Benson}, A., {et~al.} 2007, \apjs, 172, 150

\bibitem[{{Silk} \& {Mamon}(2012)}]{silk&manon2012}
{Silk}, J. \& {Mamon}, G.~A. 2012, Research in Astronomy and Astrophysics, 12,
  917

\bibitem[{{Silverman}(1986)}]{silverman1986}
{Silverman}, B.~W. 1986, {Density estimation for statistics and data analysis}

\bibitem[{{Sobral} {et~al.}(2014){Sobral}, {Best}, {Smail}, {Mobasher},
  {Stott}, \& {Nisbet}}]{sobral2014}
{Sobral}, D., {Best}, P.~N., {Smail}, I., {et~al.} 2014, \mnras, 437, 3516

\bibitem[{{Stefanon} \& {Marchesini}(2013)}]{stefanon&marchesini2013}
{Stefanon}, M. \& {Marchesini}, D. 2013, \mnras, 429, 881

\bibitem[{{Strazzullo} {et~al.}(2006){Strazzullo}, {Rosati}, {Stanford},
  {Lidman}, {Nonino}, {Demarco}, {Eisenhardt}, {Ettori}, {Mainieri}, \&
  {Toft}}]{strazzullo2006}
{Strazzullo}, V., {Rosati}, P., {Stanford}, S.~A., {et~al.} 2006, \aap, 450,
  909

\bibitem[{{Umetsu} {et~al.}(2012){Umetsu}, {Medezinski}, {Nonino}, {Merten},
  {Zitrin}, {Molino}, {Grillo}, {Carrasco}, {Donahue}, {Mahdavi}, {Coe},
  {Postman}, {Koekemoer}, {Czakon}, {Sayers}, {Mroczkowski}, {Golwala}, {Koch},
  {Lin}, {Molnar}, {Rosati}, {Balestra}, {Mercurio}, {Scodeggio}, {Biviano},
  {Anguita}, {Infante}, {Seidel}, {Sendra}, {Jouvel}, {Host}, {Lemze},
  {Broadhurst}, {Meneghetti}, {Moustakas}, {Bartelmann}, {Ben{\'{\i}}tez},
  {Bouwens}, {Bradley}, {Ford}, {Jim{\'e}nez-Teja}, {Kelson}, {Lahav},
  {Melchior}, {Moustakas}, {Ogaz}, {Seitz}, \& {Zheng}}]{umetsu2012}
{Umetsu}, K., {Medezinski}, E., {Nonino}, M., {et~al.} 2012, \apj, 755, 56

\bibitem[{{van der Burg} {et~al.}(2013){van der Burg}, {Muzzin}, {Hoekstra},
  {Lidman}, {Rettura}, {Wilson}, {Yee}, {Hildebrandt}, {Marchesini},
  {Stefanon}, {Demarco}, \& {Kuijken}}]{vanderburg2013}
{van der Burg}, R.~F.~J., {Muzzin}, A., {Hoekstra}, H., {et~al.} 2013, \aap,
  557, A15

\bibitem[{{Vulcani} {et~al.}(2011){Vulcani}, {Poggianti},
  {Arag{\'o}n-Salamanca}, {Fasano}, {Rudnick}, {Valentinuzzi}, {Dressler},
  {Bettoni}, {Cava}, {D'Onofrio}, {Fritz}, {Moretti}, {Omizzolo}, \&
  {Varela}}]{vulcani2011}
{Vulcani}, B., {Poggianti}, B.~M., {Arag{\'o}n-Salamanca}, A., {et~al.} 2011,
  \mnras, 412, 246

\bibitem[{{Vulcani} {et~al.}(2012){Vulcani}, {Poggianti}, {Fasano}, {Desai},
  {Dressler}, {Oemler}, {Calvi}, {D'Onofrio}, \& {Moretti}}]{vulcani2012}
{Vulcani}, B., {Poggianti}, B.~M., {Fasano}, G., {et~al.} 2012, \mnras, 420,
  1481

\bibitem[{{Vulcani} {et~al.}(2013){Vulcani}, {Poggianti}, {Oemler}, {Dressler},
  {Arag{\'o}n-Salamanca}, {De Lucia}, {Moretti}, {Gladders}, {Abramson}, \&
  {Halliday}}]{vulcani2013}
{Vulcani}, B., {Poggianti}, B.~M., {Oemler}, A., {et~al.} 2013, \aap, 550, A58

\bibitem[{{Whitmore} {et~al.}(1993){Whitmore}, {Gilmore}, \&
  {Jones}}]{Whitmore+93}
{Whitmore}, B.~C., {Gilmore}, D.~M., \& {Jones}, C. 1993, \apj, 407, 489

\bibitem[{{Ziparo} {et~al.}(2013){Ziparo}, {Popesso}, {Biviano}, {Finoguenov},
  {Wuyts}, {Wilman}, {Salvato}, {Tanaka}, {Ilbert}, {Nandra}, {Lutz}, {Elbaz},
  {Dickinson}, {Altieri}, {Aussel}, {Berta}, {Cimatti}, {Fadda}, {Genzel}, {Le
  Flo'ch}, {Magnelli}, {Nordon}, {Poglitsch}, {Pozzi}, {Portal}, {Tacconi},
  {Bauer}, {Brandt}, {Cappelluti}, {Cooper}, \& {Mulchaey}}]{Ziparo+13}
{Ziparo}, F., {Popesso}, P., {Biviano}, A., {et~al.} 2013, \mnras, 434, 3089

\end{thebibliography}

\end{document}